\documentclass[prd,showpacs,showkeys,superscriptaddress,nofootinbib,twocolumn,floatfix]{revtex4}
\usepackage{hyperref}
\hypersetup{colorlinks=true, linkcolor=blue, citecolor=black}
\usepackage{amsfonts}
\usepackage{graphicx,color,amsmath,amssymb}
\usepackage{subcaption}
\usepackage[activate={none},final]{microtype} 
\usepackage{parskip}  
\usepackage{braket}
\usepackage{natbib}
\usepackage{caption}
\newcommand{\qqb}{\ensuremath{Q\bar{Q}\;}}

\def\ba{\begin{eqnarray}}
\def\ea{\end{eqnarray}}
\def\be{\begin{equation}}
\def\ee{\end{equation}}

\begin{document}

\title{In-medium heavy quark-antiquark $T$-matrix without partial wave expansion}
\author{Anurag Tiwari}
\email{anurag.tiwari128@gmail.com}
\affiliation{Department of Applied Physics, Nanjing University of Science and Technology, Nanjing 210094, China}
\author{Min He}
\email{minhephys@gmail.com}
\affiliation{Department of Applied Physics, Nanjing University of Science and Technology, Nanjing 210094, China}

\date{\today}

\begin{abstract}
The T-matrix of the heavy quark-antiquark (\qqb) pair interacting through a screened Cornell potential in the quark-gluon plasma (QGP) is computed without employing the partial wave expansion. This is compared with the results obtained using the conventional method of partial wave expansion. Through a comprehensive survey over a range of screening mass and center-of-mass energy, which, when combined, determine the orbital angular momentum involved in the scattering through the associated force range and incident momentum, it is demonstrated that a substantial number of partial wave terms are necessary to precisely match the full scattering amplitude directly obtained from the method without partial wave expansion. For moderate screening masses (corresponding to one to two times the crossover transition temperature) and center-of-mass energy, the number of partial waves required for satisfactory matching turns out as large as 10-20, substantially larger than what one would expect based on simply measuring the magnitudes of the first few partial wave amplitudes. This highlights the efficiency of the new method in directly obtaining the full scattering amplitude needed for further computation of phenomenological observables. We also employ the new method to calculate the T-matrix at center-of-mass energies below the \qqb mass threshold, and demonstrate the presence of bound states of different orbital quantum numbers simultaneously in a single energy scan.
\end{abstract}


\maketitle

\section{Introduction}

\label{sec:intro} 

Ultra-relativistic high-energy collision experiments aim to create and study the quark-gluon plasma (QGP), a novel phase of nuclear matter. Quarkonia—bound states of heavy quark-antiquark pairs (\qqb, where Q is a charm or bottom quark)—serve as key probes for understanding color-force modifications in QGP \cite{Rapp:2008tf, Andronic:2015wma, Zhao:2020jqu}. As different heavy quarkonia are characterized by different sizes in vacuum, they are expected to be influenced progressively in the inverse order of their binding energies when immersed in the medium. The earliest prediction of a sequential quarkonia suppression \cite{Matsui:1986dk}, primarily due to color-charge screening of \qqb potential in QGP, has been observed experimentally in the bottomonia sector \cite{CMS:2023lfu}. However, a static screening scenario alone is insufficient; in-medium dynamical effects, such as inelastic collisions \cite{Kharzeev:1994pz, Grandchamp:2001pf}, reflecting the imaginary part of the potential \cite{Laine:2006ns}, generate state-dependent dissociation widths. At LHC energies, the enhanced production of charm quarks enables charmonium regeneration from uncorrelated $c\bar{c}$ pairs \cite{Thews:2000rj, Braun-Munzinger:2000csl, Grandchamp:2003uw}, while bottomonium regeneration occurs via in-medium transitions to different bound states of a single correlated pair \citet{Brambilla:2023hkw,Yao:2020xzw}. The regeneration rate, through the fluctuation-dissipation theorem, is related to the dissociation rate \cite{Grandchamp:2005yw} and consequently to the imaginary potential. Theoretically, the strength of screening (mass shift), the in-medium binding energies and the reaction rates are all encoded in the quarkonia spectral functions~\cite{Andronic:2024oxz}. 

The in-medium quarkonia spectral properties can be computed through the first-principle lattice QCD approach, which provides information on both the real and imaginary parts of the \qqb potential \cite{Bala:2021fkm, Bazavov:2023dci, Ali:2025iux}. Due to the heavy quark mass $M_{Q}\gg T$, the energy $k_{0}$ of the exchanged gluons between the \qqb pair is parametrically smaller than their momentum $|\vec{k}|\sim T$, implying that a static picture remains valid in the QGP medium. Therefore, it is anticipated that the predominant interactions between the \qqb pair are of potential nature, which can then be resummed by the scattering T-matrix equation. The thermodynamical T-matrix framework has been employed in the computation of spectral functions of quarkonia, heavy-light mesons and diffusion coefficient for heavy quarks in the QGP \cite{Mannarelli:2005pz, Cabrera:2006wh,vanHees:2007me, Riek:2010fk, Liu:2017qah}. Its advantages lie in providing a unified framework to analyze both the bound and the continuum states of \qqb pair in medium \cite{Cabrera:2006wh}. It is based on a three-dimensional (3D) reduction of the relativistically covariant four-dimensional Bethe-Salpeter equation. The two most common methods for the 3D reduction in literature are those of Blankenbeclar-Sugar \cite{BbS1966} and Thompson \cite{Thompson1970}. The 3D reduction enables one to solve the two-body Bethe-Salpeter scattering amplitude using an instantaneous (thus nonrelativistic) potential as input, while retaining the relativistic unitarity \cite{BbS1966,Thompson1970}. Higher order relativistic corrections can also be appropriately included in this approach \cite{Cabrera:2006wh, Riek:2010fk}. 
Lattice QCD data of quarkonia correlators and heavy quark susceptibilities, when applied to this approach, may provide first-principle constraints for the extraction of in-medium \qqb potential as well as spectral functions \cite{Riek:2010fk, Liu:2017qah}. On the other hand, direct application of the phenomenologically motivated potential in this approach, e.g. the Cornell-type potential parameterized with color-screening has also turned out useful in the T-matrix calculations of in medium heavy quark systems \cite{ He_2023, Wu:2025hlf}.

The conventional T-matrix calculations of \qqb system have employed the partial wave expansion method (termed ``PWE" in this work) \cite{Haftel:1970zz, Mannarelli:2005pz, Cabrera:2006wh,vanHees:2007me, Riek:2010fk, Liu:2017qah}, to simplify the numerical solution of the 3D reduced T-matrix equation. It is advantageous when the \qqb potential exhibits rotational symmetry. The T-matrix equation in PWE method assumes the form of a one dimensional integral equation, which can be formulated as a matrix inversion problem \cite{Haftel:1970zz}. In problems with small angular momentum, the scattering rates can be computed with first few partial wave amplitudes. Similarly, the PWE method is suitable for computing spectral functions or bound states for a specific channel with definite orbital angular momentum (e.g., s-wave or p-wave) . However, there are situations where these advantages break down. Firstly, when the \qqb potential is not rotationally symmetric, e.g. in cases where one aims to address the anisotropy of the QGP medium or if the center-of-mass momentum of \qqb pair cannot be neglected. Secondly, the full T-matrix is related to the its PWE components as (see Sec.~\ref{subsec:pwe_method} for details)
\be
\label{eqn:T_PWE_sum_intro}
T(\vec{q'},\vec{q})  = 4\pi \sum_{l = 0}^{\infty} (2l+1)T_{l}(q',q) P_{l}(\hat{q'}\cdot\hat{q}).
\ee
The convergence of the sum in the above equation is marred by the presence of the degeneracy factor $(2l+1)$ and Legendre polynomials of order `$l$'(denoted by $P_{l}(x)$). Firstly, the factor $(2l+1)$ can counter the decrease of the partial wave amplitudes $T_{l}$ as $l$ increases and thus significantly slow down the convergence of the PWE series. Furthermore, the oscillatory nature of Legendre polynomials may necessitate the inclusion of a substantial number of PWE terms to achieve precise matching with the full amplitude on the left hand side which is usually a smooth function of the involved variables and is needed for precision calculations of phenomenological observables, {\it e.g.}, heavy quark diffusion coefficient~\cite{He_2023}. The magnitude of the angular momentum involved in a problem depends upon the impact parameter and the momentum of the projectile, implying that a large number of PWE terms will be required to reproduce the full amplitude in case one of them is large. Consequently, it is worthwhile to explore the solution of T-matrix equations without employing the partial wave expansion (termed ``no-PWE" in this work). Indeed, the no-PWE method has been pursued for the nonrelativistic Lippmann-Schwinger equation addressing the nucleon-nucleon scattering in vacuum~\cite{HOLZ1988131,Rice1993, Elster_1998}. 

 In this work, we apply the no-PWE method \cite{Elster_1998} to compute the \qqb scattering mediated by a temperature dependent potential \cite{Karsch1988ER} in the QGP. We calculate the pertinent T-matrix utilizing both PWE and no-PWE methods and compare them against each other for a wide range of Debye screening masses $m_D$ and center-of-mass energies $E_{\text{CM}}$. Working with the Thompson form of 3D-reduced T-matrix equation \cite{Mannarelli:2005pz}, we demonstrate for the scattering problem (i.e., $E_{\text{CM}}$ above the \qqb mass threshold) that the number of PWE terms required to match the full amplitude obtained directly from the no-PWE method is substantially larger than one would expect from simply measuring the magnitudes of first few partial wave amplitudes, especially at small screening mass or large $E_{\text{CM}}$. Applying the no-PWE method to the bound state problem where $E_{\text{CM}}$ is below mass threshold, we demonstrate that the no-PWE T-matrix can accommodate all bound states of different orbital angular momentum simultaneously, for the first time in literature to our knowledge. For the purpose of simply demonstrating the power of the no-PWE method for computing the T-matrix, we refrain from coupling the two-body \qqb scattering equation to the one-body $Q$ self-energy equation in the present work, which would require a self-consistent iteration of the integral system and will be pursued in the future. 

The article is organized as follow. In Sec.~\ref{sec:formalism} the basics of 3D-reduced T-matrix equation is outlined. Following the brief account of the PWE method and its numerical implementation in Sec.~\ref{subsec:pwe_method}, the formalism of the no-PWE method and details of its numerical implementation are discussed in Sec.~\ref{subsec:nopwe_method} and \ref{subsec:noPWE - Numerical method}. Then in Sec.~\ref{sec:results} we move on to present in detail the numerical computations and results of our work. First we discuss in Sec.~\ref{subsec:T_l_comparison} the behavior of first few partial wave amplitudes for a range of screening masses. Next, in Sec.~\ref{subsec:PWE vs no-PWE} we discuss the comparison between the PWE and no-PWE methods for the scattering problem, by varying the screening mass $m_D$ and  $E_{\text{CM}}$ over a wide range. In Sec.~\ref{subsec:offshell_comparison} the half offshell T-matrix amplitudes are compared between the PWE and no-PWE approaches.  In the last Sec.~\ref{sec:Bound_State}, we discuss the computation of below mass threshold T-matrix (pertaining to bound state problem) in the no-PWE approach, from the perspectives of both numerical demonstration and analytical understanding. In Sec.~\ref{sec:summary} we conclude our work and present a short summary, with potential applications in future work.

\section{T-matrix formalism}
\label{sec:formalism}

The starting point of our analysis is 3D-reduced \cite{BbS1966, Thompson1970} form of the four dimensional Bethe-Salpeter equation which has been conventionally used in \qqb T-matrix calculations. We work in the center-of-mass(CM) frame setting $\vec{P}_{CM} = 0$. The 3D-reduced equation is given as 
\ba
\label{eqn: T matrix eqn2}
T(E_{\text{CM}}, \vec{q'}, \vec{q}) &=&  V( \vec{q'}, \vec{q}) + \int {\mathrm{d}^{3}k \over (2\pi)^{3}} V( \vec{q'}, \vec{k}) \nonumber \\
& & \times G_{Q\bar{Q}}(E_{\text{CM}}, k)T(E_{\text{CM}}, \vec{k}, \vec{q})
\ea
The 3D-reduction in essence puts the propagator of virtual heavy quark(antiquark) on its positive(negative) mass-shell. The 3D-reduced two body propagator is constructed in such way that the unitarity properties of the four-dimensional Bethe-Salpeter equation are preserved. The complete two-body kernel is approximated by the potential term $V(\vec{q'},\vec{q})$ which does not depend upon $E_{\text{CM}}$. Replacing the two-body interaction kernel with the potential term amounts to approximating the full scattering series with the ladder approximation. We also ignore any relativistic corrections to the potential term, as implemented in \cite{Cabrera:2006wh,Riek:2010fk}, or the spin structure of the $T$ and $V$ matrices and the resulting complications that arise thereafter.  

The propagation of heavy quark-antiquark pairs is governed by $G_{Q\bar{Q}}(E, k)$. We employ Thompson propagator \cite{Thompson1970,Mannarelli:2005pz}, for its simple form and similarity to the non-relativistic case:
\begin{equation}
\label{eqn: Thompson prop}
G_{Q\bar{Q}}(E, k) = \frac{1}{2} \frac{1}{\omega_{k} - E/2 - i\Sigma_{I}(\omega_{k}, k)} (1 - 2 n_{F}(\omega_{k}, T)).
\end{equation}
Here, $\omega_{k} = \sqrt{k^{2}+ M^{2}}+\Sigma_{R}(\omega_{k}, k)$ and $\Sigma_{R}(\omega_{k}, k)$, $\Sigma_{I}(\omega_{k}, k)$ represent the real and imaginary parts of the heavy quark (antiquark) self-energy in the medium, respectively. $n_{F}(\omega_{k}, T)$ denotes the Fermi-Dirac thermal distribution function for heavy quarks (antiquarks):
\begin{equation}
n_{F}(\omega_{k}, T) = \frac{1}{e^{\omega_{k}/ T} + 1}.
\end{equation}
The factor of $(1 - 2 n_{F}(\omega_{k}, T))$ accounts for the thermal modification of the two-body propagator $G_{Q\bar{Q}}(E, k)$. The complete many-body treatment of this problem requires solving a coupled self-energy equation for the heavy quark (antiquark), which is inserted back into Eq.~(\ref{eqn: T matrix eqn2}). The self-energy stems from interactions between heavy quarks (antiquarks) with thermal light quarks/antiquarks and gluons present in the medium. In this work, we disregard this additional modification of single heavy quark (antiquark) properties and set the real part to zero, $\Sigma_{R}(\omega_{k}, k) = 0$. We employ a constant, albeit small, imaginary part $\Sigma_{I}$ in the propagator to mitigate the delta function-like poles of the propagator. Given that our primary focus in this work is solely on the formal comparison of PWE and no-PWE methods, we defer further details regarding these additional considerations to a future work.

\subsection{\uppercase{PWE} method}
\label{subsec:pwe_method}
The two-body interaction kernel is approximated by the potential interactions $V( \vec{q'}, \vec{q})$. Assuming the potential in position space to be given by a local, scalar term, but independent of the spin of the $Q$,$\bar{Q}$ particles, the $T$ and $V$ matrices in the momentum space are expanded in terms of their partial wave components
\ba
\label{eqn: partial wave expansion of T and V}
V(\vec{q'},\vec{q}) &=& 4\pi \sum_{l = 0}^{\infty} (2l+1)V_{l}(q',q) P_{l}(\hat{q'}\cdot\hat{q}) ,\nonumber \\
T(\vec{q'},\vec{q})  &=& 4\pi \sum_{l = 0}^{\infty} (2l+1)T_{l}(q',q) P_{l}(\hat{q'}\cdot\hat{q}) ,
\ea
with the $V(\vec{q'},\vec{q})$ obtained from the Fourier transform of the position space potential
\ba
V(\vec{q'},\vec{q}) &=& \int \mathrm{d}^{3}r V(\vec{r})  e^{ i (\vec{q}-\vec{q'} )\cdot \vec{r}  } .
\ea
Here $\hat{q} = \vec{q}/|\vec{q}|$ is the unit vector, and $P_{l}(x)$ is the Legendre polynomial of order `$l$'. Integrating out the angle associated with $\vec k$ in Eq.~(\ref{eqn: T matrix eqn2}) using the orthogonality relation for the Legendre polynomials and matching the component of $P_{l}(x)$ on both sides, one obtains the T-matrix equation for its `$l$'th-component 
\ba
\label{eqn:T_matrix_partial_wave}
T_{l}(E,q',q)&=& V_{l}(q',q) +  {2 \over \pi}\int_{0}^{\infty} \mathrm{d}k \ k^{2}   V_{l}(q',k)  \nonumber \\
& &\qquad\times G(E, k)T_{l}(E, k, q),
\ea
where 
\be
\label{eqn:V_matrix_pwe}
V_{l}(q',q) = \int \mathrm{d}r\;r^{2} j_{l}(q' r) j_{l}(q r) V(r),
\ee
and $j_{l}(r)$ is the spherical Bessel function of order `$l$'.
After the partial wave expansion, the resulting T-matrix equation is solved using the algorithm of Haftel-Tabakin \cite{Haftel:1970zz}. After the discretizations of momentum variables using the Gauss-Legendre quadratures, denoted as $q_{j}$, Eq.~(\ref{eqn:T_matrix_partial_wave}) can be written in the matrix form as 
\begin{align}
[T_{l}(E)]_{ij} &= [V_{l}]_{ij} +  \frac{2}{\pi} \sum_{k} w_{k}\; q_{k}^{2} [V_{l}]_{ik} [G(E)]_{kk} [T_{l}(E)]_{kj}, \nonumber \\
[T_{l}(E)] &= (1 - [V_{l}] [G(E)])^{-1} [V_{l}],
\label{eqn:T_matrix_partial_wave_matrix_form}
\end{align}
and can be solved using standard matrix inversion routines. Throughout this work we use the notation $[\ldots]$ to represent the matrix form of $T$ or $V$ matrices.
\subsection{no-\uppercase{PWE} method}
\label{subsec:nopwe_method}
In \cite{Elster_1998} the non-relativistic Lippmann-Schwinger equation for a nucleon-nucleon system in vacuum was solved without employing the partial wave expansion. After the 3D reduction has been performed, the Thompson T-matrix equation is formally similar to the Lippmann-Schwinger equation, with main difference occurring in the two-body propagator and the thermal corrections to it, both of which in the case of $\vec{P}_{CM} = 0$ can be made independent of angular coordinates. As such, the no-PWE method of \cite{Elster_1998} can be extended to the Thompson equation. We first write the 3D-reduced T-matrix equation (Eq.~\ref{eqn: T matrix eqn2}) as 
\ba
\label{eqn:T_matrix_no_PWE_1}
T(q', q, x') &=&  V(q', q, x') \nonumber \\
& & + \int_{0}^{\infty} \frac{\mathrm{d}k\;k^2}{(2\pi)^{3}} \int_{-1}^{1} \mathrm{d}x'' \int_{0}^{2\pi} \mathrm{d}\phi'' \,V(q', k, y) \nonumber \\ 
& & \quad \times G_{Q\bar{Q}}(E,k)T(k, q, x''),
\ea
where $x' = \hat{q'}\cdot\hat{q}$, $y = \hat{q'}\cdot\hat{k}$, $x'' = \hat{k}\cdot\hat{q}$, and we have suppressed the $E$ dependence of $T(E, \vec{q'},\vec{q})$ for brevity. Setting $\hat{q} = \hat{z}$ and using the angular coordinates for the unit vectors $\hat{q'} = \hat{q'}(\theta',\phi')$, $\hat{q''} = \hat{q''}(\theta'',\phi'')$, we can express $y$ as a function of $x', x''$ as 
\be
\label{eqn:unit_vector_parametrization}
y = x'x'' + \sqrt{1-x'^2}\sqrt{1-x''^2}\cos{(\phi''-\phi')}.
\ee
\begin{figure*}[htbp]
\centering
\includegraphics[width=\textwidth, keepaspectratio]{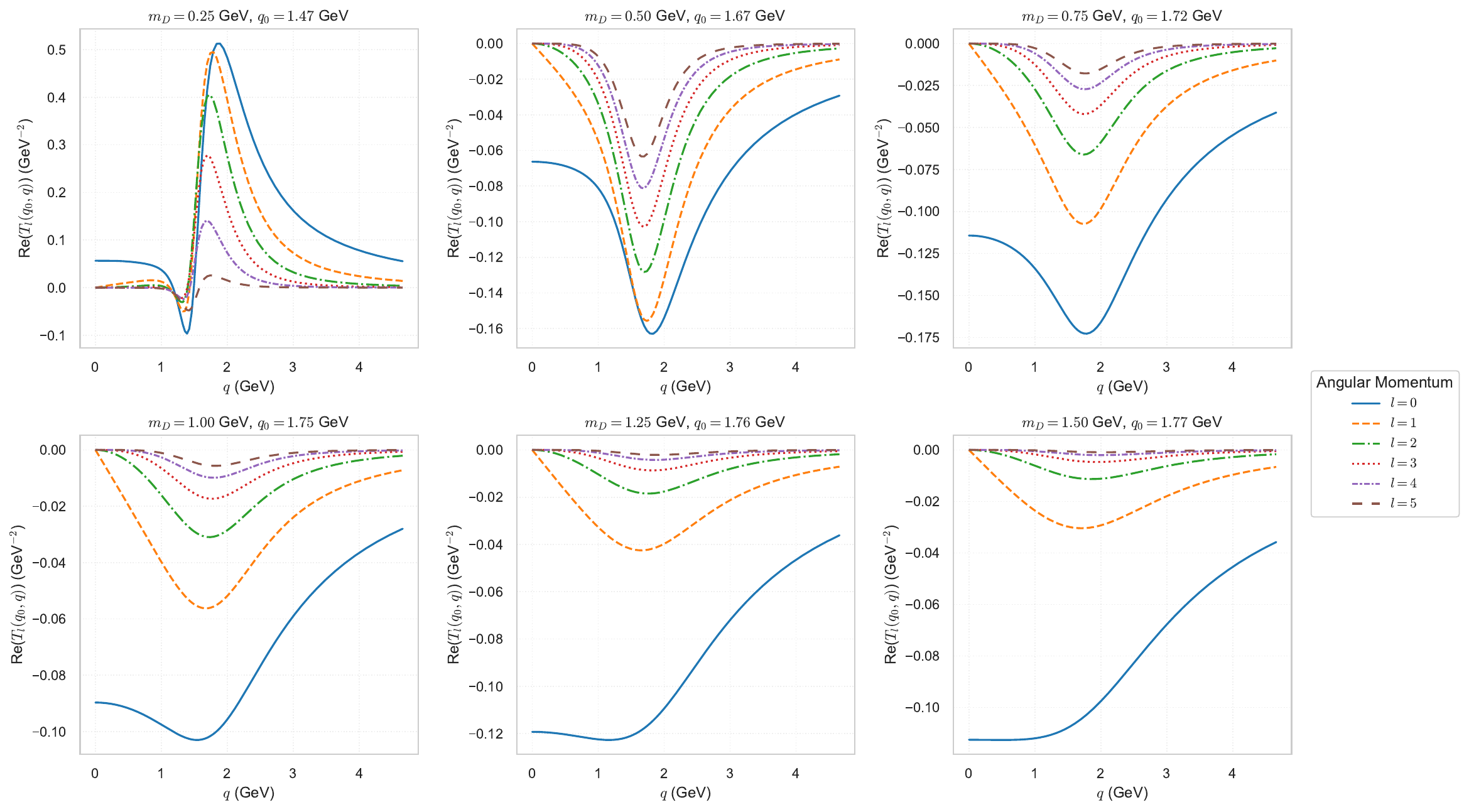}
\caption{The first five $ \text{Re}(T_{l}( q_{0}, q,))$($\text{Re}(T)$ denotes real part of the T-matrix in here and rest of the plots) as a function of `$q$' are shown for six different values of screening mass we consider. The value of onshell momentum $q_{0}$ and the corresponding screening mass $m_D$ are given in the title of the plot. We find that for all $m_D$ the $\text{Re}(T_{l})$ decreases as one goes to higher angular momentum. The decrease is however slower for the first two plots corresponding to $m_D=0.25~\mathrm{GeV}$ and $m_D=0.5~\mathrm{GeV}$.}
\label{fig:ReT2D_vs_l_all_mD}
\end{figure*}
\begin{figure*}
\centering
\includegraphics[width=\textwidth]{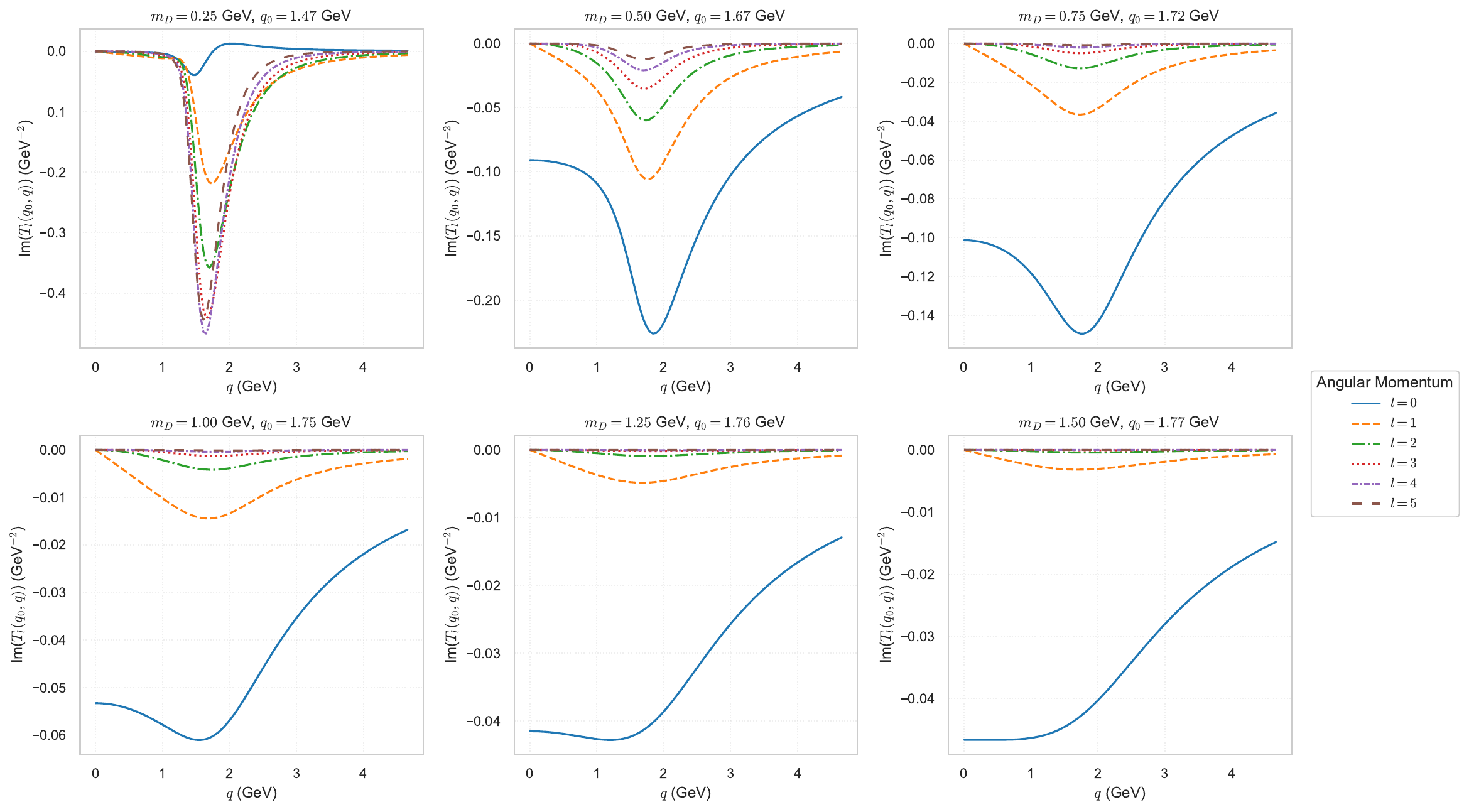}
 \caption{The first five $ \text{Im}(T_{l}( q_{0}, q,))$($\text{Im}(T)$ denotes imaginary part of the T-matrix in here and rest of the plots) as a function of `$q$' are shown for six different values of screening mass we consider. Rest same as Fig.~\ref{fig:ReT2D_vs_l_all_mD}. For the $\text{Im}(T_{l})$ the $m_D=0.25~\mathrm{GeV}$ plot stands out. Here $\text{Im}(T_{l})$ first increases as one goes to higher  `$l$', attaining its largest value for $l = 4$ after which it starts decreasing. For every other value of $m_D$ we have uniform convergence against `$l$'.}
\label{fig:ImT2D_vs_l_all_mD}
\end{figure*}
Since no term depends upon $\phi'$ we conveniently set it to zero. Substituting this back in Eq.~(\ref{eqn:T_matrix_no_PWE_1}), we get
\ba
\label{eqn:T_matrix_no_PWE_2}
& &T(q', q, x') =\frac{1}{2\pi} v(q', q, x', 1) + \nonumber \\
& & \int_{0}^{\infty} \frac{\mathrm{d}k\;k^2}{(2\pi)^{3}} \int_{-1}^{1} \mathrm{d}x'' \,v(q', k, x', x'')G_{Q\bar{Q}}(E, k)T(k, q, x''), \nonumber \\
& &
\ea
where we have introduced
\ba
\label{eq:v_small_def}
& &v(q',q,x',x'')  \nonumber \\
& &=  \int_{0}^{2\pi} \mathrm{d}\phi''  V(q',q, x'x'' + \sqrt{1-x'^2}\sqrt{1-x''^2}\cos{\phi''}). \nonumber \\
& &
\ea
In the next section we discuss how to numerically solve the no-PWE T-matrix equation.

\subsection{Numerical method for no-PWE}
\label{subsec:noPWE - Numerical method}

 After the discretizations of momentum and the angular variables, the resulting quantities in Eq.~(\ref{eqn:T_matrix_no_PWE_2}) are arrays of dimension $3$ or higher, and cannot be solved with the method of matrix inversion used in the PWE case and a different numerical method has to be implemented to solve the no-PWE T-matrix equation, which we discuss in this section. Noting that in Eq.~(\ref{eqn:T_matrix_no_PWE_2}) (remember that the CM energy dependence has been suppressed), the only variables present in the equation are $q'$ and $x'$ which through the kernel term $v(q',q,x',x'')$ couple the T-matrix on the RHS and LHS. We can utilize this by solving the equation for each $q$ independently. To understand the numerical method, first we convert Eq.~(\ref{eqn:T_matrix_no_PWE_2}) into its discrete version by using the Gauss-Legendre quadratures to convert the integration over the momentum and angular variables, $k$ and $x$ respectively, into a discrete sum. The array $T(E, q', q, x')$ for a given $E$ and $q$ can be rearranged into a vector of length $N_{q}\times N_{x}$ as $[T]_{i}$, where $N_{q}$ and $N_{x}$ are number of grid-points for the Gauss-Legendre quadratures for the momentum and the angular variables. Using the same method, the term $v(q', q, x', 1)$ can also be converted into a vector of the same length, corresponding to a fixed $q$. The integration kernel can be expressed as
\begin{figure*}[htbp] 
 \centering \includegraphics[height=0.5\textheight,width=\textwidth]{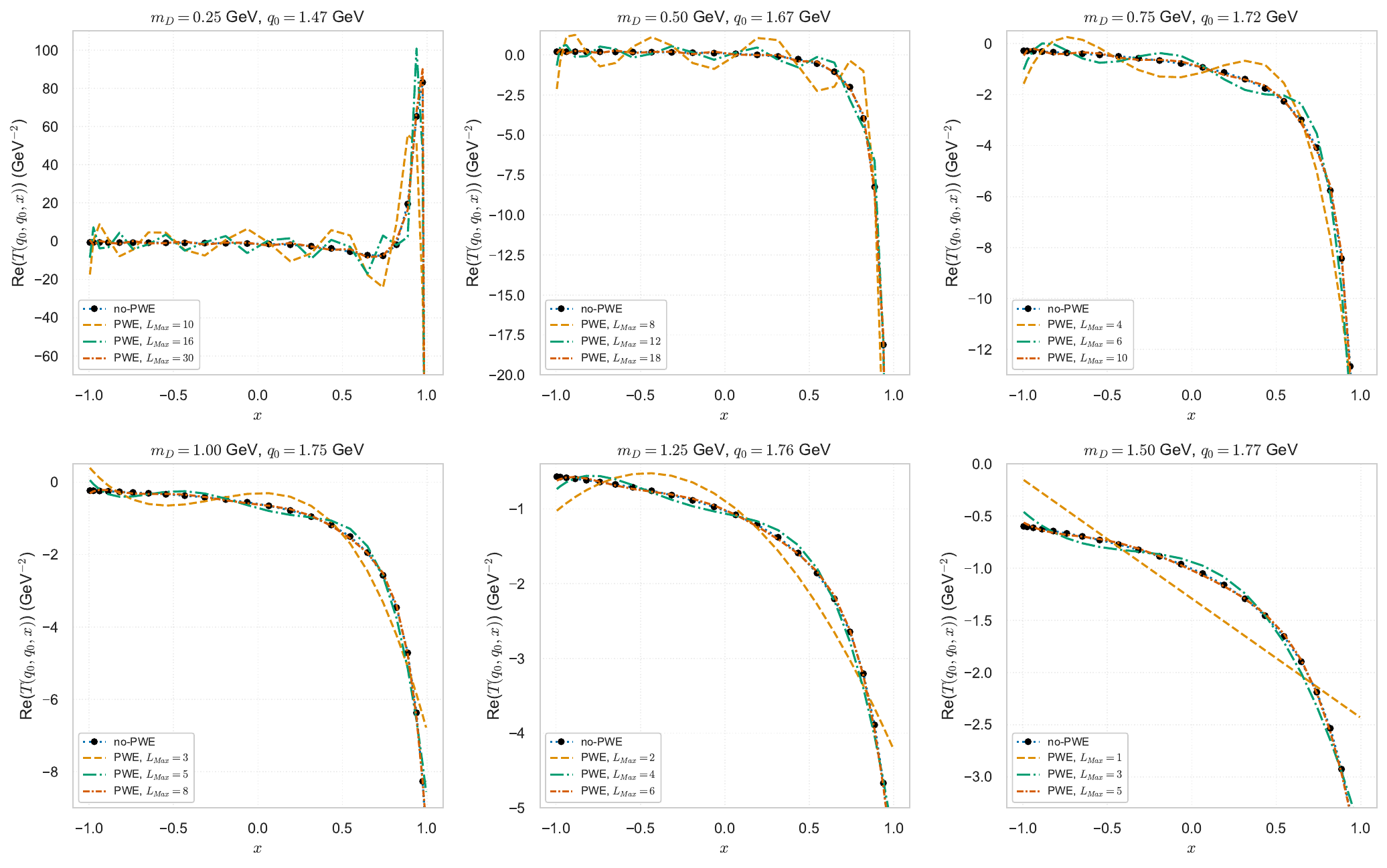}
 \caption{$\text{Re}(T(E, q_{0}, q_{0}, x))$ as a function of `$x$' is compared between the PWE and no-PWE methods, for six different values of screening mass. The labels on PWE curves denote the number of partial wave terms summed. The value of onshell momentum $q_{0}$ and the corresponding screening mass $m_D$ are given in the title of the plot. As the screening mass increases from $0.25\;\mathrm{GeV}$(left-most top row) to $1.5\;\mathrm{GeV}$(right-most bottom row), the required number of partial-wave terms decreases substantially. }
\label{fig:ReT_2Dvs3D_all_mD}
 \centering
 \includegraphics[height=0.5\textheight,width=\textwidth]{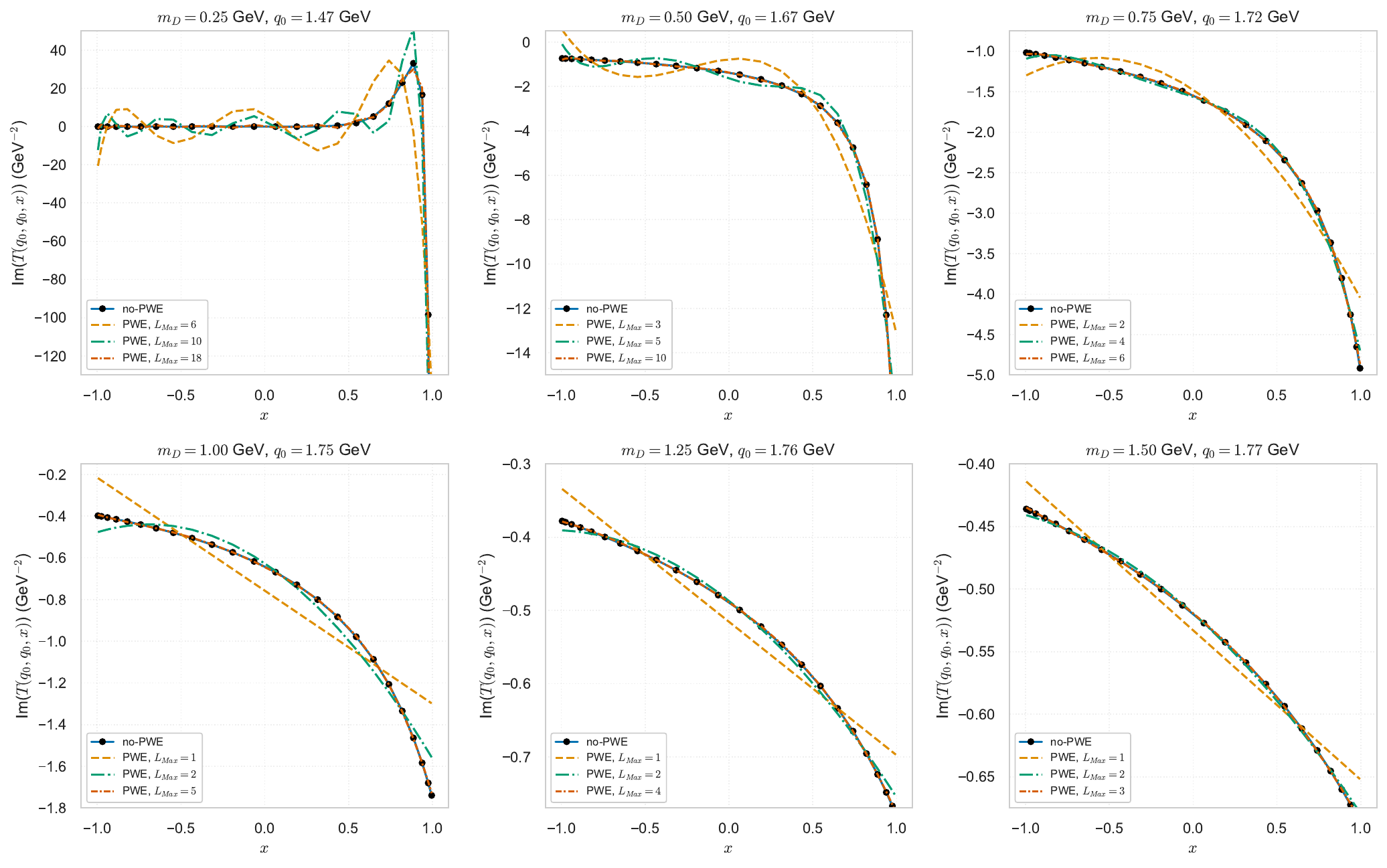}
 \caption{$\text{Im}(T(E, q_{0}, q_{0}, x))$ as a function of `$x$' is shown for six different values of screening mass we consider. Rest of the details are same as the Fig.~\ref{fig:ReT_2Dvs3D_all_mD}. One note that the required number of PWE terms to match the no-PWE result, is significant less than that for the real part of the T-matrix. }
\label{fig:ImT_2Dvs3D_all_mD}
\end{figure*}
\be
[K(E)]_{ij} = \frac{k^2_{j}}{(2\pi)^{3}}[w(k)]_{g(j)} [w(x)]_{f(j)} [v]_{ij}[G(E)]_{j},
\ee
where $[v]_{ij} = v(q'_{i}, k_{j}, x'_{i}, x''_{j})$ is a two dimensional matrix of size $(N_{q}\times N_{x})\times(N_{q}\times N_{x})$, and $[w(k)]$, $[w(x)]$ are Gauss-Legendre weights corresponding to the momentum and angular integration variables $k$ and $x''$. The function $f(i),\;g(i)$ are reminder and modulo functions respectively\footnote{Since `$i,\;j$' run over the entire range $i,j\in [1, N_{q}\times N_{x}]$, we pick the corresponding values of $[w(q)]$ and $[w(x)]$ using $i// N_{x}$(modulo operator) and $i \% N_{x}$(reminder operator), respectively.}, to assign $[w(k)]$ and $[w(x)]$ to their corresponding $q_{i}$ and $x_{j}$, as all variables have to be casted into a vector of length $N_{q}\times N_{x}$. After vectorization of all the terms in the stated manner, for a particular $q_k$ one obtains the following matrix equation
\begin{align}
\label{eq:noPWE_numerical}
[T(E)]^{k}_{i} &= \frac{1}{2\pi}[v]^{k}_{i} +  [K(E)]_{ij}[T(E)]^{k}_{j},\nonumber \\
[T(E)]^{k} &= (1 -  [K(E)])^{-1}\frac{1}{2\pi}[v]^{k}.
\end{align}
This is similar in form to the PWE matrix equation which is expected since both of them are linear equation, and can be solved using the method of matrix inversion. Then by looping over the entire range of $q_{k}$ and joining the solution $T^{k}$'s for each case, one obtains the full solution $T(q', q, x)$. The $E_{\text{CM}}$ dependence of $T(E, q', q, x)$ enters the T-matrix equation through the dependence of two-body propagator $G_{Q\bar{Q}}(E, k)$ on $E_{\text{CM}}$, and can be addressed independently of other variables present in Eq.~(\ref{eqn:T_matrix_no_PWE_2}).
Before moving to the next section, we note that we scale the Gauss-Legendre quadratures $u$'s, obtained in the range $(0, 1)$, for the momentum integration range $(0, \infty)$ using the transformation
\be
\label{eq:GL_scaling}
q = b \tan{(\frac{u\pi}{2})},
\ee
where the constant $b$ is a scaling parameter. We use $b = 0.275\;\mathrm{GeV}$, which was determined to be the optimal value that proves to align the PWE and no-PWE results efficiently across the wide range of screening masses considered in our work.

\section{Scattering State Problem: $E_{\text{CM}} > 2 M_{c}$}
\label{sec:results}
We first discuss the \qqb scattering problem, characterized by $E_{\text{CM}} > 2 M_{c}$, where $M_{c}$ is the mass of open charm quark in the QGP medium. The magnitude of angular momentum in a scattering problem depends upon two factors, the impact parameter and the incoming momentum of the projectile. The impact parameter in the \qqb scattering problem is determined by the effective range of the \qqb potential $V(r, T)$, whereas the incoming momentum in the problem depends upon the CM energy $E_{\text{CM}}$. Several  works \cite{Bala_2020, Bala:2021fkm} have addressed the determination of $V(r, T)$ on lattice, and have found that the potential exhibit screening which increases with increasing medium temperature. To model the function $V(r, T)$, we employ a Cornell type potential
\be
\label{eqn: Cornell potential}
V(r, T) = {-\alpha \over r}e^{- m_D(T)r} + {\sigma \over m_D(T)}(1- e^{- m_D(T)r} ),
\ee
which is characterized by a screening mass $m_D(T)$, and rest of the parameters are same as those used in \cite{Karsch1988ER}, $\alpha = 0.471,\;\sigma = 0.192\;\mathrm{GeV}^{2}$. The temperature dependence of $V(r, T)$ is solely contained in the screening mass $m_D(T)$, and we can vary the effective interaction range or the impact parameter in the scattering problem by varying $m_D(T)$.

The bare mass of charm quark is set to $M^{0}_{c} = 1.32 \ \mathrm{GeV}$. The function $V(r, T)$ employed in Eq.~(\ref{eqn: Cornell potential}) tends to a finite value $\sigma / m_D(T)$ as $r \rightarrow \infty$. Its Fourier transform $V(\vec{q'},\vec{q})$ exists only if $V(r, T)$ goes to zero as $r \rightarrow \infty$, therefore we subtract the $V(r = \infty, T) = \sigma/m_D(T)$ part from the potential function and add this to the heavy quark bare mass (following the line of work in \cite{Cabrera:2006wh,Riek:2010fk}), as a temperature dependent contribution to $M_{Q}(T)$ arising due to its interaction with the thermal medium, $M_{c} = M^{0}_{c} + \sigma/2m_D$. This also affects the threshold value $E_{\text{th}} = 2 M_{c}(T)$ below which bound states are formed, as $2M_{c}(T)$ decreases with increasing temperature or $m_D(T)$.
\subsection{Partial wave amplitudes $T_{l}$ }
\label{subsec:T_l_comparison}
In this section we first discuss partial wave amplitudes $T_{l}$ and their convergence, for a range of screening masses which are later employed in the comparison between the PWE and no-PWE methods. As the QGP medium expands and cools down, the \qqb pair undergoes scattering at different temperatures, and $m_D(T)$ changes throughout the medium evolution. To model this we consider six different screening masses $m_D = \{0.25, 0.5, 0.75, 1, 1.25, 1.5\}$ (GeV). Using a perturbative estimate for the screening mass $m_D\sim gT$ with $g\sim 2$, our considered range cover a moderate range of temperature $T_{c}-3\;T_{c}$, where $T_{c}\sim160\;\mathrm{MeV}$ is the pseudo-critical temperature denoting the crossover temperature for the QCD. The center of mass energy is fixed to $E_{\text{CM}} = 4.5\;\mathrm{GeV}$ for the results presented in this section. 

In a physical problem only the onshell amplitude $T_{l}(q_0, q_0)$ is needed, where $q_0$ is the onshell momentum in the scattering problem given by the onshell condition $q_0=\sqrt{E_{\text{CM}}^{2}/4 - M_{c}^{2}}$. To obtain the onshell amplitude the T-matrix Eq.~(\ref{eqn:T_matrix_partial_wave}) needs to be solved at least for the half-offshell case, $T_{l}(q_0, q)$, therefore we work with half-offshell partial wave amplitudes in this section. The comparison of first six half offshell $T_{l}(q_{0},q)$ as a function of `$q$' is presented in Fig.~\ref{fig:ReT2D_vs_l_all_mD} for the real part, and in Fig.~\ref{fig:ImT2D_vs_l_all_mD} for the imaginary part. In the figure, though $E_{\text{CM}} = 4.5\;\mathrm{GeV}$ is fixed, the onshell momentum changes slightly as the charm quark mass depends upon $m_D$ through the relation $M_{c} = M^{0}_{c}+\frac{\sigma}{2m_D}$. 

The overall observation is that both the real and imaginary part of the partial wave amplitudes reduce as $l$ increases. For the real part of $T_{l}$,  except for the two lowest screening masses, corresponding to $m_D = 0.25$ and $\;0.5\;\mathrm{GeV}$, at which the half offshell $\text{Re}T_{l}(q_{0},q)$ decreases rather slowly with increasing $l$, for the rest of the screening mass $m_{D} = 0.75-1.5\;\mathrm{GeV}$ the real part of $T_{l}(q_{0},q)$ all shows an order of magnitude decrease from $l=0$ to $l=5$. It is interesting to note that resummation of the strong attractive potential at the lowest screening mass $m_D = 0.25$\,GeV results in a significant positive $\text{Re}T_{l}$ for $q>q_0$, which disappears for weaker potentials at higher screening masses. For a fixed $l$, the magnitude of $\text{Re}T_{l}$ generally becomes smaller as $m_D$ increases. This can be understood from the fact that, when the screening mass increases, the effective range of interaction potential and thus the orbital angular momentum involved in the scattering decreases, and consequently higher partial wave amplitudes corresponding to larger $l$ values must become smaller. The imaginary part of the half offshell $T_{l}(q_{0}, q)$ at $m_D = 0.25\;\mathrm{GeV}$, first increases with $l$ up to $l=4$, and then starts decreasing as one goes to higher $l$'s (in the figure only up to $l=5$ terms are plotted but higher terms become smaller as $l$ keeps increasing). For the rest of the screening masses $m_{D} = 0.5-1.5\;\mathrm{GeV}$ the imaginary part all diminishes even faster than the real part with increasing $l$, and the $l\geq3$ terms are already barely visible  for $m_{D}\geq0.5$\,GeV. 

Based on the above observation of the diminishing behavior of these first few partial wave amplitudes with increasing angular momentum, one would expect that the first three to five partial wave amplitudes should be enough to be summed over to arrive at the full amplitude needed for further precision calculation of observables. However we will see in the next section that substantially larger number of PWE terms are necessary for reproducing the full T-matrix amplitudes, even at large screening masses $m_D=0.75, \;1\;\mathrm{GeV}$.

\subsection{PWE vs no-PWE: onshell amplitudes}
\label{subsec:PWE vs no-PWE}
The partial wave T-matrix amplitudes are related to the full T-matrix (or the no-PWE T-matrix) as
\be
\label{eqn:T_PWE_sum}
T(q', q, x) = 4\pi \sum_{l=0}^{L_{\text{Max}}} (2l+1)P_{l}(x)T_{l}(q',q),
\ee
where $L_{\text{Max}}$ is the minimum number of partial wave terms required to reproduce the no-PWE full T-matrix. In Eq.~(\ref{eqn:T_PWE_sum}), the degeneracy of the $l$-th channel is given by the factor of $(2l+1)$, which increases with increasing $l$ and acts to counter the decrease in the partial wave amplitudes $T_{l}$'s, therefore slowing down the convergence of the partial wave expansion in Eq.~(\ref{eqn:T_PWE_sum}). Further, Legendre polynomials put another angle-dependent weight for each $T_l$, which oscillate for all values of $x$ except $x=1$ ($P_l(x=1)=1$). Therefore the sum of different partial wave amplitudes does not monotonically converges to the no-PWE (full) T-matrix but rather approaches it in an oscillatory manner. These two factors combined render the overall convergence of the sum in Eq.~(\ref{eqn:T_PWE_sum}) substantially slower than what the fast decay of the magnitude of the first few partial wave amplitudes (cf. Fig.~\ref{fig:ReT2D_vs_l_all_mD} and Fig.~\ref{fig:ImT2D_vs_l_all_mD}) would suggest otherwise.

To demonstrate this, in this section we compare the T-matrix computed in the PWE and no-PWE methods, and determine the value of $L_{\text{Max}}$ in Eq.~(\ref{eqn:T_PWE_sum}) for a wide range of screening masses and center-of-mass energies. We demonstrate that the number of PWE terms required in Eq.~(\ref{eqn:T_PWE_sum}) for the onshell amplitude can get as large as $l\sim10-20$ for our considered range of screening masses and CM energies. 

In the first case we fix the CM energy to $E_{\text{CM}} = 4.5\;\mathrm{GeV}$, and vary the screening mass in \qqb potential over the same range $m_{D} = 0.25-1.5\;\mathrm{GeV}$ as done in Fig.~\ref{fig:ReT2D_vs_l_all_mD} and Fig.~\ref{fig:ImT2D_vs_l_all_mD} , thus probing the effect of varying the impact parameter (thus orbital angular momentum) in the scattering problem. The value of onshell momenta considered in our work vary around $q_{0}\sim 1.5\;\mathrm{GeV}$ which is the typical value of near-thermalized charm quark momentum, as simulated in the phenomenological transport models~\cite{He_2023}.
The comparison of the diagonal component of the onshell T-matrix amplitudes, $T(q_{0}, q_{0}, x)$, as a function of $x$ for PWE and no-PWE methods for six different values of $m_D$, are presented in Figs.~\ref{fig:ReT_2Dvs3D_all_mD} and \ref{fig:ImT_2Dvs3D_all_mD} for the real and the imaginary part, respectively. At any screening mass $m_D$, the convergence is substantially slower for the PWE T-matrix sum in Eq.~(\ref{eqn:T_PWE_sum}) than suggested by the apparent magnitudes of the first few partial wave terms $T_{l}$'s (cf. Fig.~\ref{fig:ReT2D_vs_l_all_mD} and Fig.~\ref{fig:ImT2D_vs_l_all_mD}). For a precise agreement between the two methods, up to $l=30$ PWE terms are necessary for the lowest value of $m_D=0.25~\mathrm{GeV}$. For $m_D=0.5~\mathrm{GeV}$ and $m_D=0.75~\mathrm{GeV}$ that correspond to most phenomenologically relevant QGP temperature range of $T\sim 200$-$400$\,MeV, at least $l=18$ and $l=10$ 
PWE terms are needed, respectively, for a precise matching of the real part. This number drops to $l=5$ for the largest $m_D = 1.5~\mathrm{GeV}$ we consider.

\begin{figure*}[htbp]
 \centering
 \includegraphics[height=0.5\textheight, width=\textwidth]{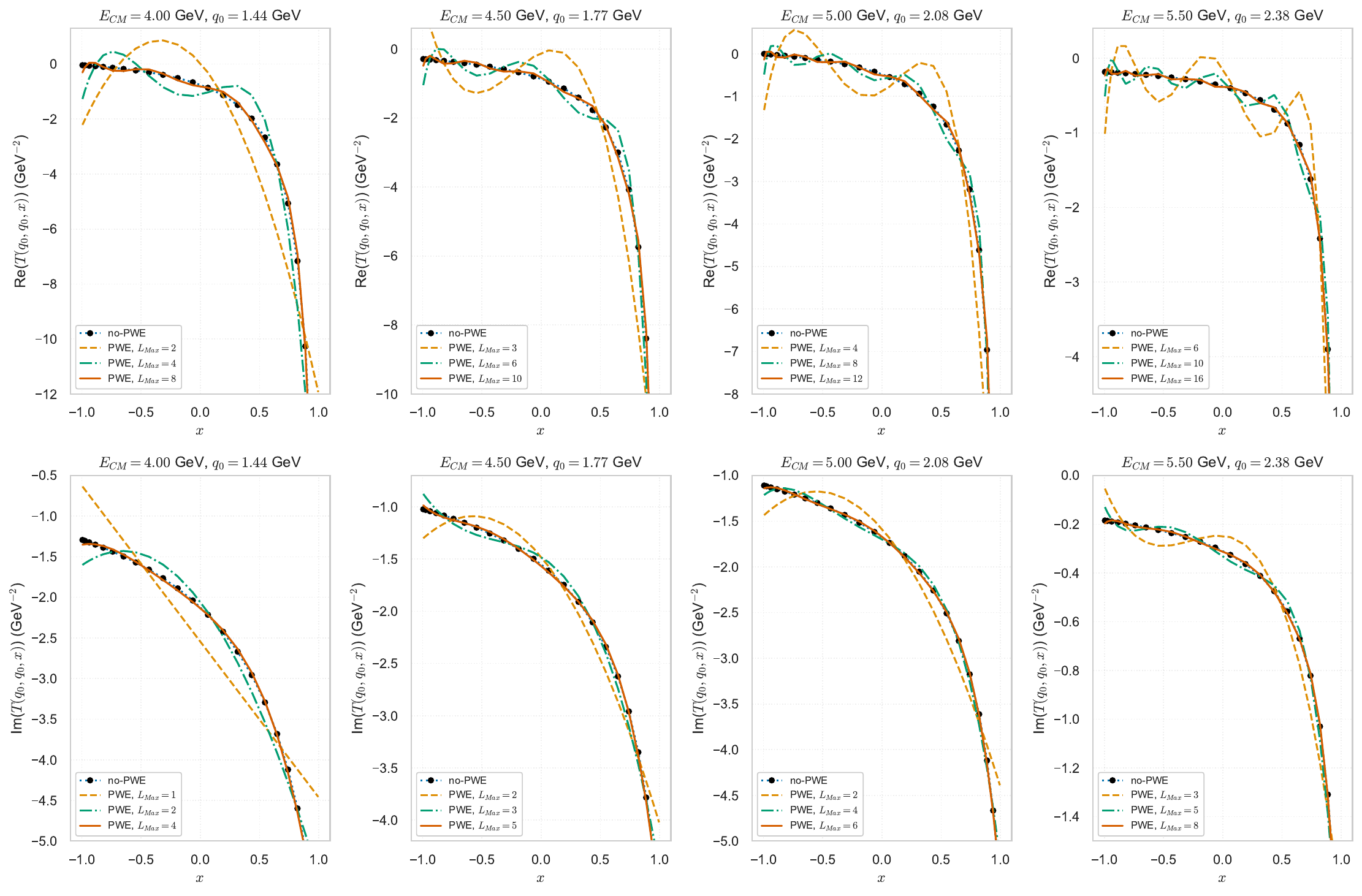}
 \caption{The $\text{Re}(T(E, q_{0}, q_{0}, x))$(top row) and the $\text{Im}(T(E, q_{0}, q_{0}, x))$(bottom row) as a function of `$x$', is shown for four different values of $E_{\text{CM}}$. Values of $E_{\text{CM}}$ and the corresponding onshell momentum $q_{0}$ are given in the title. As one increase the $E_{\text{CM}}$, higher number of PWE terms are required to match the no-PWE result.}
\label{fig:ReT_2Dvs3D_all_Eng}  
\end{figure*}
Conversely, we find that the number of partial wave terms needed to match the imaginary part of T-matrix between PWE and no-PWE methods is substantially less than that for the real part, in general. For $m_D=0.25~\mathrm{GeV}$, only $l=18$ terms are needed, for $m_D=0.5~\mathrm{GeV}$ only $l=10$ and for $m_D=0.75~\mathrm{GeV}$ only $l=6$, PWE terms are required for a precise matching with the no-PWE results, as compared to $l=30$, $l=18$ and $l=10$ terms for the real part, respectively. To understand this, we resort to the phase shift analysis in the scattering theory, where the partial wave amplitudes can be computed as a function of phase shift $\delta_l$ for the `$l$'th channel. The real part of the T-matrix is proportional to $\cos{(\delta_{l})}\sin{(\delta_{l})}$ while the imaginary part is proportional to  $\sin^{2}{(\delta_{l})}$~\cite{Elster_1998}. The phase shift $\delta_l$ becomes small as one goes to higher `$l$', so that the real part becomes approximately proportional to $\delta_{l}$ while the imaginary part to $\delta^{2}_{l}$ appearing at a higher order. Due to this, the imaginary part in general requires less partial wave terms to match a no-PWE calculation than the real part of T-matrix.

In the second case, we change the onshell momentum in our problem, given by $q_{0} = \sqrt{E_{\text{CM}}^{2}/4 - M_{c}^{2}}$, which corresponds to changing the orbital angular momentum involved in the scattering by changing the incoming momentum of the projectile, while keeping the impact parameter fixed by employing a fixed screening mass $m_D = 0.75\;\mathrm{GeV}$. We work with four different values of $E_{\text{CM}} = 4,\; 4.5,\; 5,\; 5.5$ GeV. For our calculation this covers a range of $1.4 - 2.3\;\mathrm{GeV}$ for the onshell momentum $q_0$. The comparison of real and imaginary parts of the T-matrix computed in the PWE and no-PWE methods, for four different values of $E_{\text{CM}}$, are presented in Fig.~\ref{fig:ReT_2Dvs3D_all_Eng}. We observe that at $E_{\text{CM}}=4\;\mathrm{GeV}$ only $l=8$ terms are required for a precise matching of the real parts between the two methods, which increases to $l=10$ for $E_{\text{CM}}= 4.5\;\mathrm{GeV}$, $l=12$ for $E_{\text{CM}}= 5\;\mathrm{GeV}$ and finally to $l=16$ for $E_{\text{CM}}= 5.5\;\mathrm{GeV}$, the largest $E_{\text{CM}}$ we consider. In contrast, for a precise matching of the imaginary part of T-matrix in the two approaches, $l=4$ terms are sufficient for $E_{\text{CM}} = 4\;\mathrm{GeV}$, increasing to $l=5$ for $E_{\text{CM}} = 4.5\;\mathrm{GeV}$, $l=6$ for $E_{\text{CM}} = 5\;\mathrm{GeV}$, and finally $l=8$ for $E_{\text{CM}} = 5.5\;\mathrm{GeV}$, all significantly less than those numbers for the real part. These calculations demonstrate that, for a moderate increase in the center-of-mass energy from $E_{\text{CM}} = 4\;\mathrm{GeV}$ to 5.5\,GeV, the number of PWE terms required for a precise matching to the full amplitude doubles for both the real and the imaginary part, reiterating the conclusion drawn from the first case study that the convergence of the PWE series is not as fast as one would expect based on simply looking at the magnitudes of first few terms.

\begin{figure*}[htbp]
 \centering
 \includegraphics[width=\textwidth, keepaspectratio]{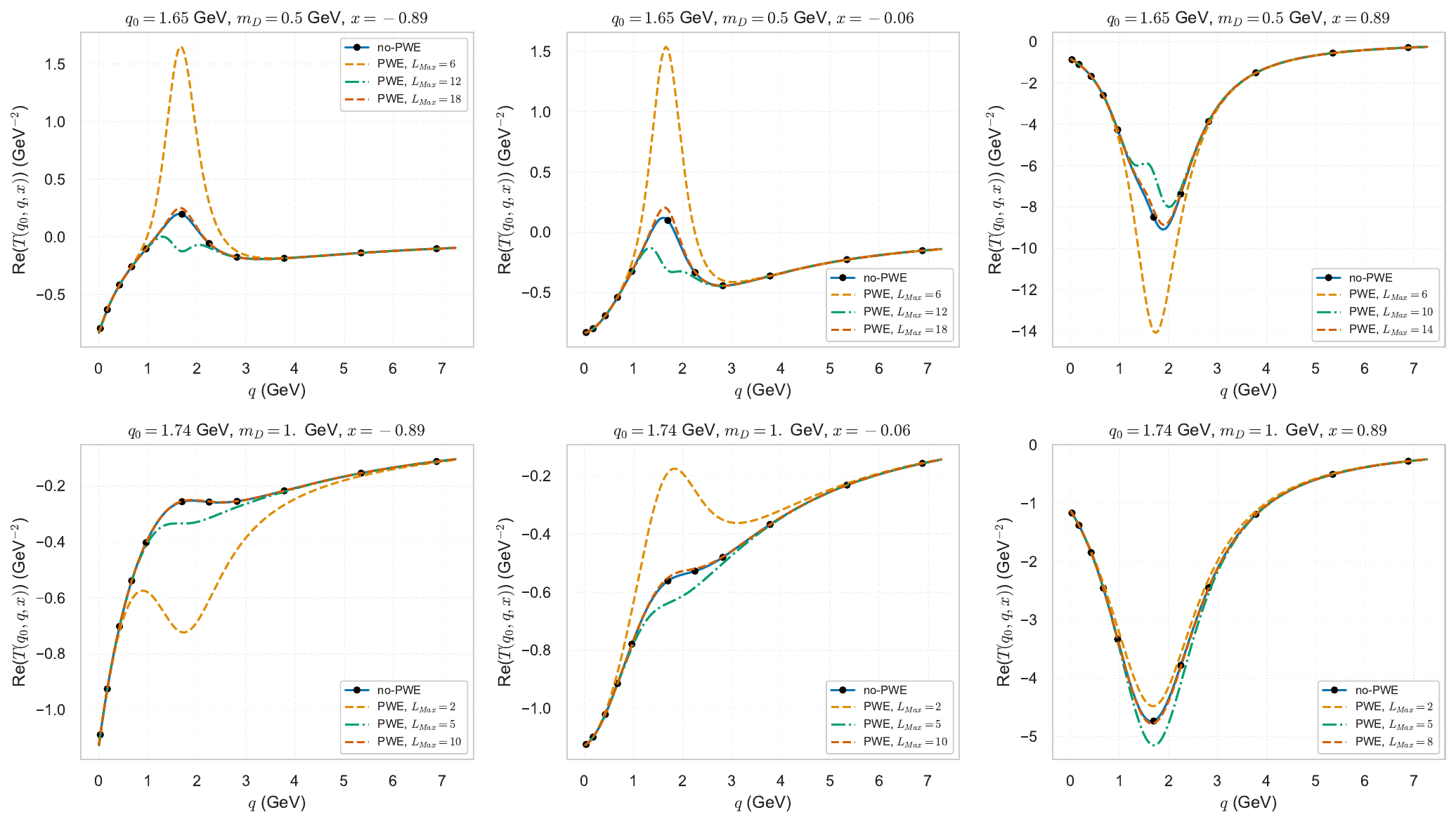}
 \caption{The real part of the half offshell $\text{Re}(T(E, q_{0}, q, x))$ as a function of `$q$', is compared between the PWE and no-PEW approach, for three different values of $x$, corresponding to the screening mass $m_D = 0.5
\;\mathrm{GeV}$ (top row) and $m_D = 1
\;\mathrm{GeV}$ (bottom row).The center-of-mass energy is fixed at $E_{\text{CM}} = 4.5\;\mathrm{GeV}$, and the values of onshell momentum $q_{0}$ and $x$ are given in the title of the plots. The half offshell curves require larger number of PWE terms to be summed near the onshell momentum. For $q\gg q_{0}$ or $q\ll q_{0}$ the lowest partial wave amplitudes dominate and the no-PWE curve can be matched with substantially less number of PWE terms.}
\label{fig:HalfOffShell_ReT_all_x}
 \centering
 \includegraphics[width=\textwidth, keepaspectratio]{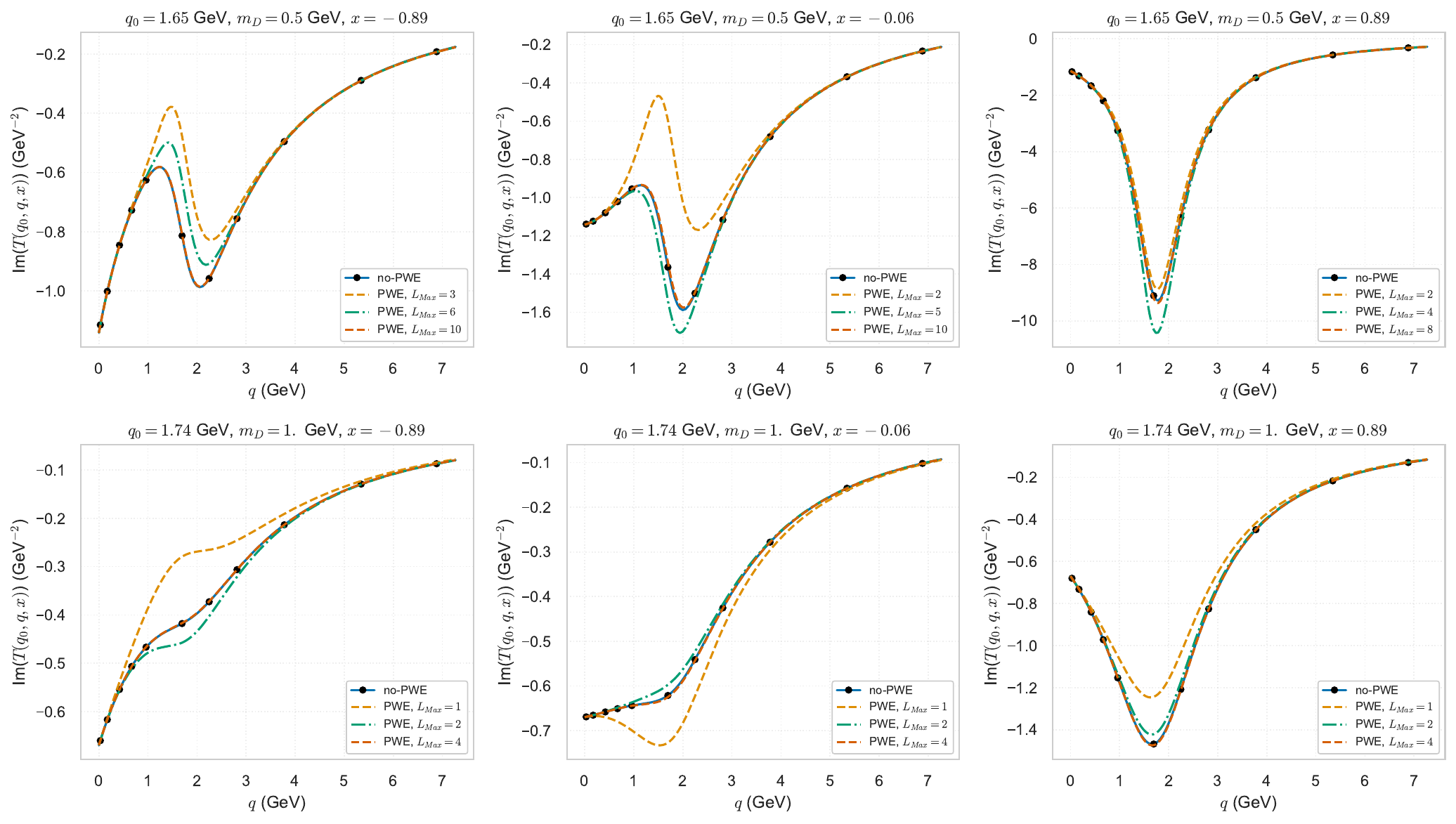}
 \caption{The imaginary part of the half offshell $\text{Im}(T(E, q_{0}, q, x))$ as a function of `$q$', is compared between the PWE and no-PEW approach for three different values of $x$, corresponding to the screening mass $m_D = 0.5
\;\mathrm{GeV}$ (top row) and $m_D = 1
\;\mathrm{GeV}$ (bottom row). Rest of the parameters are same as described in the figure above. As for the real part, $\text{Im}(T(E, q_{0}, q, x))$ also has a very small $q$ region for forward $x\approx 1$, where the largest $L_{\text{Max}}$ is needed for a precise matching with the no-PWE curve.}
\label{fig:HalfOffShell_ImT_all_x}
\end{figure*}
\subsection{PWE vs no-PWE: half offshell amplitudes}
\label{subsec:offshell_comparison}
One needs to solve for at least the half offshell T-matrix for obtaining the physical onshell amplitudes, therefore in this section we discuss the convergence of half offshell partial wave amplitudes against the no-PWE full amplitude. In Figs.~\ref{fig:HalfOffShell_ReT_all_x} and \ref{fig:HalfOffShell_ImT_all_x} the real and imaginary parts of the half offshell T-matrix amplitudes $T(q_0, q, x)$ as a function of offshell momentum $q$ are compared between the PWE and no-PWE methods, for three different values of $x$ and  two different screening masses $m_D = 0.5\;\mathrm{GeV}$ and $m_D = 1\;\mathrm{GeV}$. The center-of-mass energy is fixed at $E_{\text{CM}} = 4.5\;\mathrm{GeV}$. The value of corresponding onshell momenta are given in the title of the plots.

One observes that for both the real and imaginary parts of these half offshell amplitudes, when the offshell momentum $q$ is far away from the onshell momentum $q_0$, the required numbers of PWE terms for a precise matching to the no-PWE results are substantially less than in the onshell region, characterized by $q\approx q_{0}$, with a spread of $\sim 1$-$2\;\mathrm{GeV}$ in the offshell momentum. This can be understood by observing the behavior of the half offshell partial wave amplitudes plotted in Figs.~\ref{fig:ReT2D_vs_l_all_mD} and \ref{fig:ImT2D_vs_l_all_mD}. Near the onshell region $q\approx q_{0}$, the partial wave amplitudes $\text{Re}T_{l}(q_0, q)$ and  $\text{Im}T_{l}(q_0, q)$ achieve their maximum strengths. Furthermore, the decrease of these half offshell partial wave amplitudes with increasing $l$ is much slower in the onshell region, when compared to the offshell region $q\gg q_{0}$ or $q\ll q_{0}$ where the lowest few $\text{Re}T_{l}(q_0, q)$ and $\text{Im}T_{l}(q_0, q)$ for $l\sim0-2$ possess much larger strengths than higher partial waves and thus dominate the PWE sum in Eq.~(\ref{eqn:T_PWE_sum}).
Near the onshell region, the slowly decreasing (when $l$ increases) partial wave amplitudes are further augmented by the factor $2l+1$, and the Legendre polynomials $P_{l}(x)$ which introduce oscillations, making the convergence of the PWE sum even slower.

When increasing $m_D$ from $0.5\;\mathrm{GeV}$ to $1\;\mathrm{GeV}$, the number ($L_{\text{Max}}$) of these half offshell PWE terms required for a good matching to the no-PWE amplitude also becomes significantly smaller, same as found for the onshell amplitudes in Figs.~\ref{fig:ReT_2Dvs3D_all_mD}-\ref{fig:ReT_2Dvs3D_all_Eng}. But what's more interesting here is the $x$-dependence of the $L_{\text{max}}$ for these half offshell amplitudes. For a given $m_D$, at $x=0.89$ which is close to the forward direction ($\theta_{\vec{q'}\vec{q}}\sim0$ and thus $x\sim 1$), the magnitude of both the real and imaginary part of the full no-PWE amplitude is much larger than at other $x=-0.89$ or $-0.06$ (as already seen from the $x$-dependence of the onshell amplitudes in  Figs.~\ref{fig:ReT_2Dvs3D_all_mD}-\ref{fig:ReT_2Dvs3D_all_Eng}). This is because at the forward angle with $P_{l}(x\sim1)\sim1$ for any $l$, different partial wave amplitudes add up constructively through Eq.~(\ref{eqn:T_PWE_sum}), in contrast to the destructive sum at other angles due to oscillations arising from the Legendre polynomials. For the same reason, at $x=0.89$ the width of the near-onshell region over which a substantial number of PWE terms through Eq.~(\ref{eqn:T_PWE_sum}) are needed for a precise matching to the no-PWE results appears to be narrower and the corresponding $L_{\text{Max}}$ turns out smaller, compared to the cases at the other two angles.

\section{ Bound State Problem: $E_{\text{CM}}< 2 M_{c}$}
\label{sec:Bound_State}
When the center-of-mass energy $E_{\text{CM}}$ falls below the mass threshold $2 M_{c}$, the solution of Thomson equation (Eqs.~\ref{eqn: T matrix eqn2} and \ref{eqn: Thompson prop}) is anticipated to exhibit bound state poles when $E_{\text{CM}}$ approaches $ E_{nl}$, where $E_{nl}<2 M_{c}$ represents the bound state energy of the \qqb eigenstate characterized by its principal and orbital quantum numbers, $n$ and $l$, respectively. In the conventional PWE approach, each PWE component $T_l$ has to be solved in a separate integral equation  (Eq.~\ref{eqn:T_matrix_partial_wave}), therefore one needs to scan the energy range separately for each distinct orbital quantum number $l$. For $E_{\text{CM}}<2 M_{c}$ the ``onshell" momentum becomes imaginary, while the T-matrix equation is defined and solved in terms of real momenta. Several works in literature \cite{Mannarelli:2005pz, Cabrera:2006wh} have suggested an analytical continuation of the T-matrix from the imaginary momentum axis to $q'=0,\; q=0$ and have employed the zero-zero component $T(E_{\text{CM}}, 0, 0)$ to demonstrate the bound state poles, manifesting as peaks in the real and the imaginary parts. Though this method works for demonstrating the S-wave pole, for bound states of higher orbital quantum numbers, $l>0$, one faces problem as the seed potential term $V_{l}(q', q) \underset{q’, q \to 0}{\rightarrow} 0 $ (see Eq.~(\ref{eqn:V_matrix_pwe})), therefore suppressing the peak structure in the T-matrix. To overcome this hurdle, the authors in \cite{Cabrera:2006wh} plotted the determinant of the transition matrix $(1 - [V_{l}] [G(E)])$ which is zero at the pole location, to demonstrate the P-wave bound states, though this method also relies on a separate scan of the energy range for bound states of distinct orbital quantum numbers.  

\begin{figure*}[htbp]
 \centering 
 \includegraphics[width=\textwidth, keepaspectratio]{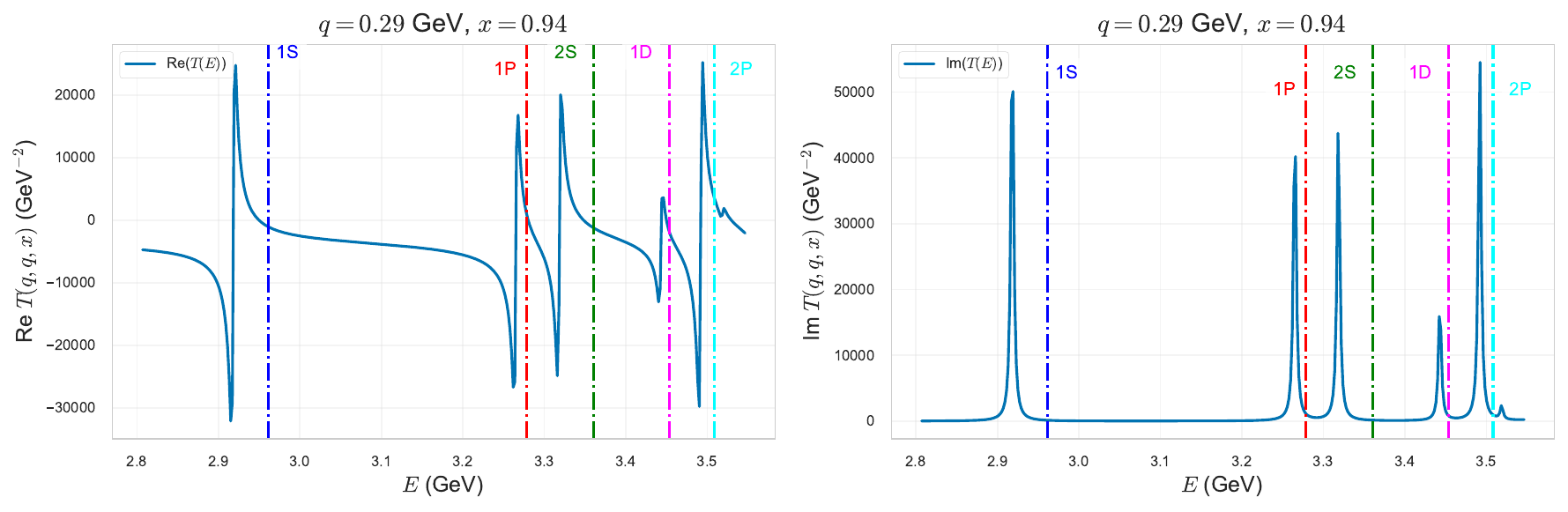}
 \caption{The $\text{Re}(T(E, q, q, x))$ (left) and $\text{Im}(T(E, q, q, x))$(right) part of \qqb T-matrix obtained from solving the Thompson-equation with a small screening mass $m_D = 0.18\;\mathrm{GeV}$, are plotted as a function of $E_{\text{CM}}$. The values of `$q$' and `$x$' given in the title of the plot are determined such that the residue strength of the bound states of distinct orbital and principal quantum numbers, is large enough to render them visible, simultaneously. The bound state masses obtained from solving the Schr\"{o}dinger equation are plotted as vertical lines and labeled with their quantum numbers, serving as a reference for the pole masses obtained from the Thompson equation. }
\label{fig:bound_state_Thomprop}
\end{figure*}

The no-PWE T-matrix, in principle, contains contributions from all the bound states of different orbital quantum numbers, which implies that a single energy scan is sufficient to demonstrate their presence or absence. In spirit of the onshell plots in Figs.~\ref{fig:ReT_2Dvs3D_all_mD}-\ref{fig:ReT_2Dvs3D_all_Eng}, we plot the real and imaginary parts of the diagonal term $T(E, q, q, x)$ as a function of CM energy $E$ in  Fig.~\ref{fig:bound_state_Thomprop}, for a small screening mass $m_D = 0.18\;\mathrm{GeV}$ in the potential for which $V(r, T)$ resemble the vacuum potential, demonstrating peak structures corresponding to multiple bound states of distinct principal and orbital quantum numbers in a single energy scan, for the first time in literature to our knowledge. In Fig.~\ref{fig:bound_state_Thomprop}, results from solving the non-relativistic Schr\"{o}dinger equation for the \qqb pair are also shown, to ascertain the quantum numbers of the bound state poles obtained in the Thompson equation. The T-matrix poles exhibit a small but noticeable shift downward compared to the eigenvalues obtained from the Schr\"{o}dinger equation. The real and imaginary parts of the T-matrix around pole energies exhibit peak structures even in the offshell momentum of the T-matrix, and P or D wave states can be demonstrated in a similar manner to the S-wave state, something which is not possible with the $T_{l}(E_{\text{CM}}, 0, 0)$ component~\cite{Mannarelli:2005pz}. 

To understand why this is possible in principle, we scrutinize the non-relativistic Lippmann-Schwinger equation which is formally similar to the 3D reduced Thompson equation, the main difference lying in the two-body propagators (Green's function) which in the Lippmann-Schwinger case contain the full non-relativistic Hamiltonian operator $H$, facilitating derivation of analytical results without the complications present in the relativistic Thompson equation. Thus by analyzing the former, one can gain insights into the behavior of residue at the pole locations for the later. The Lippmann-Schwinger equation can be written in the form 
\be
\label{eqn:Lippmann_Schwinger}
T =  V + V G_{NR} V,
\ee
where $G_{NR}(E,\vec{k}) = (E - H + i \Sigma_{I})^{-1}$, $V$ is the non-relativistic potential operator and $H = 2 M_{c}+k^{2}/M+ V$ is the non-relativistic Hamiltonian operator of the \qqb system, excluding the center-of-mass kinetic energy term.
\begin{figure*}[htbp]
 \centering 
 \includegraphics[width=\textwidth, keepaspectratio]{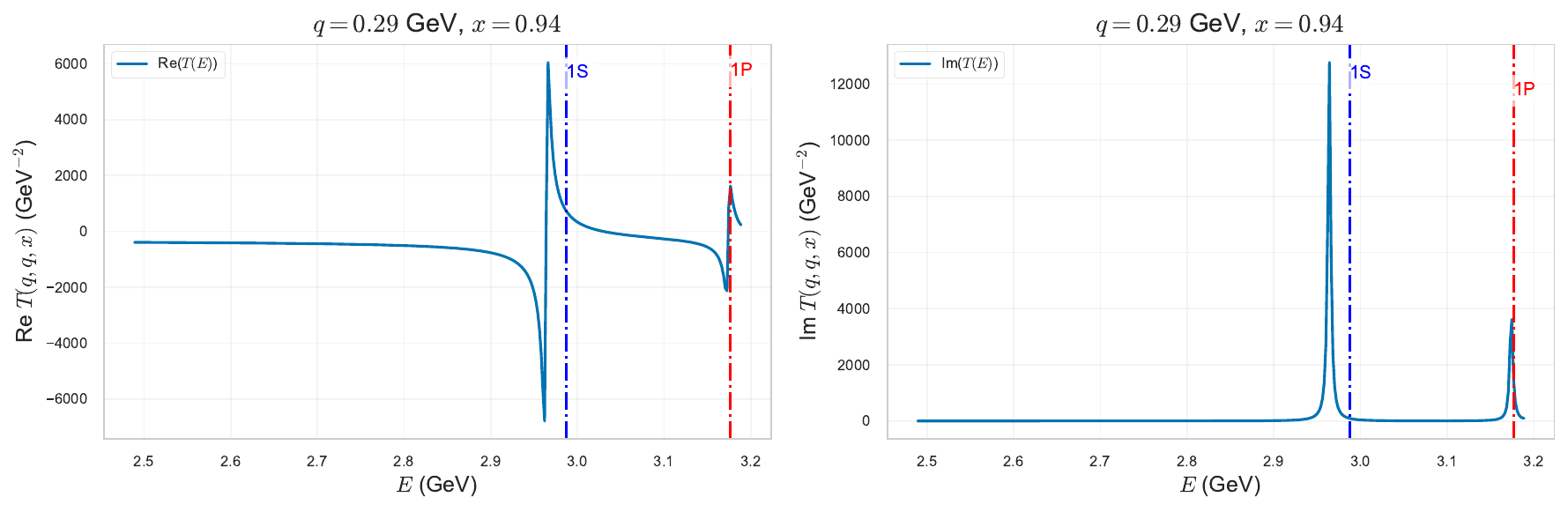}
 \caption{The results from solving the Thompson-equation for $\text{Re}(T(E, q, q, x))$ (left) and $\text{Im}(T(E, q, q, x))$(right) are plotted as a function of $E_{\text{CM}}$ for the screening mass $m_D = 0.35\;\mathrm{GeV}$. The values of `$q$' and `$x$' given in the title are same as those used in Fig.~\ref{fig:bound_state_Thomprop}. With the increase in screening mass is, higher excited states melt away, leaving only the $1S$ and $1P$ states. The vertical lines are bound state masses obtained from solving the Schr\"{o}dinger equation.}
\label{fig:bound_state_Thomprop_mu035}
\end{figure*}
 Inserting complete set of states of $H$,
\be
\mathbb{I} = \sum_{nl}\ket{\psi_{nl}}\bra{\psi_{nl}} + \int_{\alpha}\ket{\psi_{\alpha}}\bra{\psi_{\alpha}},
\nonumber
\ee
where $\ket{\psi_{nl}}$ and $\ket{\psi_{\alpha}}$ represent the bound and scattering states of $H$, respectively, and considering the T-matrix in the momentum eigenstate basis $\ket{\vec{q}},\ket{\vec{q'}}$, we obtain
\begin{align}
\label{eq: T_bound_state_1}
T(E, \vec{q'}, \vec{q}) & = \bra{\vec{q'}} V \ket{\vec{q}} + \sum_{nl}  \frac{\bra{\vec{q'}} V \ket{\psi_{nl}} \bra{\psi_{nl}}V\ket{\vec{q}}}{E - E_{nl} + i\Sigma_{I}} \nonumber  \\
& \qquad +\int_{\alpha} \frac{\bra{\vec{q'}} V \ket{\psi_{\alpha}} \bra{\psi_{\alpha}}V\ket{\vec{q}}}{E - E_{\alpha} + i\Sigma_{I}} .
\end{align}
For scattering states we have $E_{\alpha}>2M_{c}$ while $E_{nl}<2M_{c}$ for the bound states, therefore when the CM energy $E$ is below mass threshold in Eq.~(\ref{eq: T_bound_state_1}), one can make a distinction between the pole terms containing the factor $(E - E_{nl} + i\Sigma_{I})^{-1}$ from other terms which do not contain any singularity, and Eq.~(\ref{eq: T_bound_state_1}) can be separated into two parts as 
\begin{align}
\label{eq: T_bound_state_2}
T(E, \vec{q'}, \vec{q}) & = \sum_{nl}  \frac{\bra{\vec{q'}} V \ket{\psi_{nl}} \bra{\psi_{nl}}V\ket{\vec{q}}}{E - E_{nl} + i\Sigma_{I}} + \text{(regular terms)}
\end{align}
We can express the scalar product containing the potential operator as
\begin{align}
 \bra{\vec{q}} V \ket{\psi_{nl}} & = \int\mathrm{d}^{3}q_{1} \int\mathrm{d}^{3}r \braket{\vec{q}|\vec{r}}\bra{\vec{r}} V\ket{\vec{q}_{1}}\braket{\vec{q}_{1}|\psi_{nl}} \nonumber\\
 & = \int\mathrm{d}^{3}q_{1} \int\mathrm{d}^{3}r \frac{e^{i(\vec{q}\cdot\vec{r} -\vec{q}_{1}\cdot\vec{r})}}{(2\pi)^3} V(r) \psi_{nl}(q_{1})Y_{lm}(\hat{q}_{1}),
\end{align}
which after expanding the exponential factors in terms of spherical harmonics
\begin{equation}
e^{i\vec{q}\cdot\vec{r}} = 4\pi\sum_{lm} i^{l} j_{l}(q r)Y_{lm}(\hat{q})Y^{*}_{lm}(\hat{r})
\end{equation}
and doing the angular integrations, yields
\begin{align}
\label{eq: define_glq}
   \bra{\vec{q}} V \ket{\psi_{nl}} & =  \int\mathrm{d}q_{1}\;q_{1}^{2} V_{l}(q, q_{1}) \psi_{nl}(q_{1}) Y_{l m}(\hat{q}) \nonumber \\
   & = g_{l}(q) Y_{l m}(\hat{q}),
\end{align}
where $g_{l}(q)$ is defined as
\be
 g_{l}(q) = \int\mathrm{d}q'\;q'^{2}\frac{2}{\pi}V_{l}(q, q') \psi_{nl}(q'),
\ee
and $V_{l}(q', q)$ was defined in Eq.~(\ref{eqn:V_matrix_pwe}).
Finally, using the identity 
\begin{equation}
    P_{l}(\hat{q}'\cdot\hat{q}) = \frac{4\pi}{2l+1}\sum_{m}   Y_{l m}(\hat{q}')  Y^{*}_{l m}(\hat{q}) 
\end{equation}
and Eq.~(\ref{eq: define_glq}), Eq.~(\ref{eq: T_bound_state_2}) can be written as 
\begin{equation}
\label{eq: T_near_Enl_2}
  T(E, \vec{q'}, \vec{q}) = \sum_{nl} \frac{R_{l}(q', q , x)}{E - E_{nl} + i\Sigma_{I}} + (\text{regular terms})
\end{equation}
where $x = \hat{q}'\cdot\hat{q}$ and
\begin{align}
\label{eq:g_l_q}
R_{l}(q', q, x) &= \frac{2l+1}{4\pi} g_{l}(q')g_{l}(q) P_{l}(x).
\end{align}
Using Eq.~(\ref{eq: T_near_Enl_2}) we see that when $E$ is close to a pole energy $E_{nl}$, the regular terms are negligible and T-matrix is dominated by the pole term containing the residue $R_{l}(q', q, x)$, which for the real $\text{Re}(T)$ and imaginary part $\text{Im}(T)$ of the T-matrix (ignoring the regular terms) are given by
\begin{align}
\label{eq:ReT_and_ImT_residues_near_boundstates}
\text{Re}(T) &= \sum_{nl}   \frac{(E - E_{nl}) R_{l}(q', q, x)}{(E - E_{nl})^2 + \Sigma_{I}^{2}}, \nonumber \\
\text{Im}(T) &= \sum_{nl}   \frac{\Sigma_{I} R_{l}(q', q, x)}{(E - E_{nl})^2 + \Sigma_{I}^{2}}.
\end{align}
Eq.~(\ref{eq:ReT_and_ImT_residues_near_boundstates}) demonstrate that at the pole positions both real and imaginary parts of the T-matrix have a peak for even offshell values of $q$ and $q'$ and the pole locations are independent of the momenta $q',\;q$ and the angular variable $x$, as only the residue strength depends upon these variables. For a fixed $q',\;q$ and $x$, the real part changes its sign around the pole position, while the imaginary part exhibit a peak which vanishes at the non-pole energies. Since the residue term $R_{l}(q', q , x)$ depends upon the product of functions $g_{l}(q)$, $g_{l}(q')$ and $P_{l}(x)$, for the angular variable $x$ we find that using $x\approx 1$ is best suited for demonstrating the presence of peak structures in different $l$-states, as $P_{l}(x\rightarrow1)\rightarrow 1$ for all $l$ whereas for other values of $x$, $P_{l}(x)$ oscillates between positive and negative values affecting the residue term, and makes the peak structure less visible for some states. The suitable value of $q$ is determined as such for which the function $g_{l}(q)$ is not small for any quantum numbers $nl$ under consideration. Therefore an optimal combination of the variables $q$ and $x$ can be selected to ensure that the residue strength at all pole locations is significant enough, for rendering the peak structures visible for all bound states simultaneously. This provides explanation for our results in Figs.~\ref{fig:bound_state_Thomprop}, \ref{fig:bound_state_Thomprop_mu035}, and for the values of $q$ and $x$ used in there. 

The binding energy of the \qqb bound states decreases with an increase in the QGP temperature, or equivalently the screening mass $m_D$, which is the basis of the melting picture originally formulated in \cite{Matsui:1986dk}. In Fig.~\ref{fig:bound_state_Thomprop_mu035} we have plotted the real and imaginary parts of the T-matrix for a larger screening mass $m_D = 0.35\;\mathrm{GeV}$. As the screening mass increases, the excited states $2S$, $1D$ and $2P$ have melted and only $1S$ and $1P$, two lowest lying bound states survive. The binding energies of the \qqb bound states are related to their pole mass as $E_{\psi}(T) = 2 M_{Q}(T) - M_{\psi}(T)$, where $M_{\psi}(T)$ is the pole mass of the state labeled as $\psi$ at temperature $T$. The binding energies decrease significantly from $E_{1S} = 0.794\;\mathrm{GeV}$, $E_{1P} = 0.442\;\mathrm{GeV}$ for the screening mass $m_D = 0.18\;\mathrm{GeV}$, to  $E_{1S} = 0.223\;\mathrm{GeV}$, $E_{1P} = 0.0126\;\mathrm{GeV}$ when the screening mass is increased to $0.35\;\mathrm{GeV}$, due to the weakening of \qqb potential at larger screening mass.   
\section{Summary}
\label{sec:summary}
A comprehensive comparison between the conventional partial wave expansion method and a new method without partial wave expansion is conducted for solving the \qqb scattering T-matrix equation in the QGP. By comparing the new method against the PWE method for a wide range of screening masses and center-of-mass energies, which, when combined, determine the orbital angular momentum in the scattering problem, it was demonstrated that the number of partial wave amplitudes needed for a precise matching with the full no-PWE amplitude is substantially larger than what one would expect from measuring the magnitude of the first few partial wave amplitudes. The presence of the degeneracy factor $2l+1$ and Legendre polynomial $P_{l}(x)$ corresponding to the `$l$'-th angular momentum channel in the partial wave sum, significantly slows down the convergence of PWE amplitudes to the full no-PWE amplitude.

We first discussed the leading half offshell partial wave amplitudes, and found both the real and imaginary parts exhibit a monotonic decrease with increasing $l$, except for the two lowest screening masses for which the decrease is rather slow or not monotonic. By comparing the onshell scattering amplitudes calculated in the PWE and the no-PWE methods, we demonstrated that the minimum number `$L_{\text{max}}$' of partial wave amplitudes required for a precise matching between the two methods, decreases monotonically as the screening of \qqb potential becomes stronger, or the center-of-mass energy is lowered. For phenomenologically relevant values of screening masses and center-of-mass energies, $L_{\text{max}}$ can reach as high as $l\sim 10-20$. The comparison of half offshell amplitudes calculated in the two methods, demonstrates that the near-onshell region requires larger number of the PWE terms for a precise matching, while in the far offshell region significantly lower number of PWE terms are sufficient. These calculations suggest that for a precision calculation of \qqb observables which requires calculating the full in-medium scattering amplitude (i.e., T-matrix), the no-PWE method may be more suited and efficient than the conventional PWE method. In particular, the oscillations introduced by the Legendre polynomials throughout the intermediate values of $x$ when summing over PWE amplitudes of different $l$'s to approximate the full $T$ matrix can lead to persistent fluctuations around the genuine value and thus cause significant uncertainties.
  
In calculations of below mass threshold T-matrix pertaining to the \qqb bound states, first working with a small screening mass $m_D = 0.18\;\mathrm{GeV}$, which resembles a vacuum potential, we demonstrated that the no-PWE method can be employed for determination of all bound states of distinct orbital quantum numbers in a single energy scan, which is impossible in the partial wave $T$-matrix equation. Using the non-relativistic Lippmann-Schwinger equation as a proxy that allows for convenient analytical derivations, we showed that the bound states peak structures in the real and imaginary parts of T-matrix are present in the whole offshell domain, given the momentum and angular coordinate are selected in a manner that the residue strength at pole locations are sufficiently large for all bound states simultaneously. The bound state peaks of all but the $1S$ and $1P$ states, melt away when the screening mass is increased to $m_D = 0.35\;\mathrm{GeV}$ and the binding energies of the two remaining bound states also reduce significantly.

There are several directions along which our present work can be further developed. First, calculations of heavy quarkonium spectral functions whether on the lattice or from the T-matrix approach have been thus far restricted to vanishing center-of-mass momentum. The conventional PWE method of the T-matrix approach may be further complicated by the presence of another momentum vector in this connection. On the contrary, the no-PWE method is expected to be more suited for this purpose, since in the ``collinear" frame with total (center-of-mass) momentum $P=(P^0,0,0,P^3)$, the two-particle propagator $G_{\qqb}$ can still be made independent of the intermediate azimuthal angle $\phi''$~\cite{Caia:2003ke}. It is the latter fact that renders the T-matrix equation tractable without partial wave expansion (c.f. Eqs.~(\ref{eqn:T_matrix_no_PWE_1}, \ref{eqn:unit_vector_parametrization}) in Sec.~\ref{subsec:nopwe_method}). In particular, in order to perform a complete calculation, we need to solve the coupled integral system of both two-body scattering T-matrix equation and the one-body single Q self-energy equation, which has been neglected in the present work but ensures a self-consistent many-body treatment. Secondly, the T-matrix equation can be extended to the case of heavy quark-light quark/antiquark/gluon scattering, based on the realization that energy transfer between the heavy quark and light QGP constituent is still parametrically small compared to the 3-momentum transfer and thus the interaction remains of potential type~\cite{Riek:2010fk}. For this, the application of the no-PWE method may allow for a more direct and precise computation of the full scattering amplitude, which can then be used to evaluate the heavy quark transport coefficient. We defer these considerations for future works.
\begin{acknowledgments}
This work was supported by the National Natural Science Foundation of China (NSFC) under Grant No. 12475141 and the Fundamental Research Funds for the Central Universities No. 30925020109.
\end{acknowledgments}
\bibliography{references}

\begin{thebibliography}{34}
\expandafter\ifx\csname natexlab\endcsname\relax\def\natexlab#1{#1}\fi
\expandafter\ifx\csname bibnamefont\endcsname\relax
  \def\bibnamefont#1{#1}\fi
\expandafter\ifx\csname bibfnamefont\endcsname\relax
  \def\bibfnamefont#1{#1}\fi
\expandafter\ifx\csname citenamefont\endcsname\relax
  \def\citenamefont#1{#1}\fi
\expandafter\ifx\csname url\endcsname\relax
  \def\url#1{\texttt{#1}}\fi
\expandafter\ifx\csname urlprefix\endcsname\relax\def\urlprefix{URL }\fi
\providecommand{\bibinfo}[2]{#2}
\providecommand{\eprint}[2][]{\url{#2}}

\bibitem[{\citenamefont{Rapp et~al.}(2010)\citenamefont{Rapp, Blaschke, and
  Crochet}}]{Rapp:2008tf}
\bibinfo{author}{\bibfnamefont{R.}~\bibnamefont{Rapp}},
  \bibinfo{author}{\bibfnamefont{D.}~\bibnamefont{Blaschke}}, \bibnamefont{and}
  \bibinfo{author}{\bibfnamefont{P.}~\bibnamefont{Crochet}},
  \bibinfo{journal}{Prog. Part. Nucl. Phys.} \textbf{\bibinfo{volume}{65}},
  \bibinfo{pages}{209} (\bibinfo{year}{2010}).

\bibitem[{\citenamefont{Andronic et~al.}(2016)}]{Andronic:2015wma}
\bibinfo{author}{\bibfnamefont{A.}~\bibnamefont{Andronic}}
  \bibnamefont{et~al.}, \bibinfo{journal}{Eur. Phys. J. C}
  \textbf{\bibinfo{volume}{76}}, \bibinfo{pages}{107} (\bibinfo{year}{2016}).

\bibitem[{\citenamefont{Zhao et~al.}(2020)\citenamefont{Zhao, Zhou, Chen, and
  Zhuang}}]{Zhao:2020jqu}
\bibinfo{author}{\bibfnamefont{J.}~\bibnamefont{Zhao}},
  \bibinfo{author}{\bibfnamefont{K.}~\bibnamefont{Zhou}},
  \bibinfo{author}{\bibfnamefont{S.}~\bibnamefont{Chen}}, \bibnamefont{and}
  \bibinfo{author}{\bibfnamefont{P.}~\bibnamefont{Zhuang}},
  \bibinfo{journal}{Prog. Part. Nucl. Phys.} \textbf{\bibinfo{volume}{114}},
  \bibinfo{pages}{103801} (\bibinfo{year}{2020}).

\bibitem[{\citenamefont{Matsui and Satz}(1986)}]{Matsui:1986dk}
\bibinfo{author}{\bibfnamefont{T.}~\bibnamefont{Matsui}} \bibnamefont{and}
  \bibinfo{author}{\bibfnamefont{H.}~\bibnamefont{Satz}},
  \bibinfo{journal}{Phys. Lett. B} \textbf{\bibinfo{volume}{178}},
  \bibinfo{pages}{416} (\bibinfo{year}{1986}).

\bibitem[{\citenamefont{Tumasyan et~al.}(2024)}]{CMS:2023lfu}
\bibinfo{author}{\bibfnamefont{A.}~\bibnamefont{Tumasyan}} \bibnamefont{et~al.}
  (\bibinfo{collaboration}{CMS}), \bibinfo{journal}{Phys. Rev. Lett.}
  \textbf{\bibinfo{volume}{133}}, \bibinfo{pages}{022302}
  (\bibinfo{year}{2024}).

\bibitem[{\citenamefont{Kharzeev and Satz}(1994)}]{Kharzeev:1994pz}
\bibinfo{author}{\bibfnamefont{D.}~\bibnamefont{Kharzeev}} \bibnamefont{and}
  \bibinfo{author}{\bibfnamefont{H.}~\bibnamefont{Satz}},
  \bibinfo{journal}{Phys. Lett. B} \textbf{\bibinfo{volume}{334}},
  \bibinfo{pages}{155} (\bibinfo{year}{1994}).

\bibitem[{\citenamefont{Grandchamp and Rapp}(2001)}]{Grandchamp:2001pf}
\bibinfo{author}{\bibfnamefont{L.}~\bibnamefont{Grandchamp}} \bibnamefont{and}
  \bibinfo{author}{\bibfnamefont{R.}~\bibnamefont{Rapp}},
  \bibinfo{journal}{Phys. Lett. B} \textbf{\bibinfo{volume}{523}},
  \bibinfo{pages}{60} (\bibinfo{year}{2001}).

\bibitem[{\citenamefont{Laine et~al.}(2007)\citenamefont{Laine, Philipsen,
  Romatschke, and Tassler}}]{Laine:2006ns}
\bibinfo{author}{\bibfnamefont{M.}~\bibnamefont{Laine}},
  \bibinfo{author}{\bibfnamefont{O.}~\bibnamefont{Philipsen}},
  \bibinfo{author}{\bibfnamefont{P.}~\bibnamefont{Romatschke}},
  \bibnamefont{and} \bibinfo{author}{\bibfnamefont{M.}~\bibnamefont{Tassler}},
  \bibinfo{journal}{JHEP} \textbf{\bibinfo{volume}{03}}, \bibinfo{pages}{054}
  (\bibinfo{year}{2007}).

\bibitem[{\citenamefont{Thews et~al.}(2001)\citenamefont{Thews, Schroedter, and
  Rafelski}}]{Thews:2000rj}
\bibinfo{author}{\bibfnamefont{R.~L.} \bibnamefont{Thews}},
  \bibinfo{author}{\bibfnamefont{M.}~\bibnamefont{Schroedter}},
  \bibnamefont{and} \bibinfo{author}{\bibfnamefont{J.}~\bibnamefont{Rafelski}},
  \bibinfo{journal}{Phys. Rev. C} \textbf{\bibinfo{volume}{63}},
  \bibinfo{pages}{054905} (\bibinfo{year}{2001}).

\bibitem[{\citenamefont{Braun-Munzinger and
  Stachel}(2000)}]{Braun-Munzinger:2000csl}
\bibinfo{author}{\bibfnamefont{P.}~\bibnamefont{Braun-Munzinger}}
  \bibnamefont{and} \bibinfo{author}{\bibfnamefont{J.}~\bibnamefont{Stachel}},
  \bibinfo{journal}{Phys. Lett. B} \textbf{\bibinfo{volume}{490}},
  \bibinfo{pages}{196} (\bibinfo{year}{2000}).

\bibitem[{\citenamefont{Grandchamp et~al.}(2004)\citenamefont{Grandchamp, Rapp,
  and Brown}}]{Grandchamp:2003uw}
\bibinfo{author}{\bibfnamefont{L.}~\bibnamefont{Grandchamp}},
  \bibinfo{author}{\bibfnamefont{R.}~\bibnamefont{Rapp}}, \bibnamefont{and}
  \bibinfo{author}{\bibfnamefont{G.~E.} \bibnamefont{Brown}},
  \bibinfo{journal}{Phys. Rev. Lett.} \textbf{\bibinfo{volume}{92}},
  \bibinfo{pages}{212301} (\bibinfo{year}{2004}).

\bibitem[{\citenamefont{Brambilla~et al.}(2023)}]{Brambilla:2023hkw}
\bibinfo{author}{\bibfnamefont{N.}~\bibnamefont{Brambilla~et al.}},
  \bibinfo{journal}{Phys. Rev. D} \textbf{\bibinfo{volume}{108}},
  \bibinfo{pages}{L011502} (\bibinfo{year}{2023}).

\bibitem[{\citenamefont{Yao et~al.}(2021)\citenamefont{Yao, Ke, Xu, Bass, and
  M{\"u}ller}}]{Yao:2020xzw}
\bibinfo{author}{\bibfnamefont{X.}~\bibnamefont{Yao}},
  \bibinfo{author}{\bibfnamefont{W.}~\bibnamefont{Ke}},
  \bibinfo{author}{\bibfnamefont{Y.}~\bibnamefont{Xu}},
  \bibinfo{author}{\bibfnamefont{S.~A.} \bibnamefont{Bass}}, \bibnamefont{and}
  \bibinfo{author}{\bibfnamefont{B.}~\bibnamefont{M{\"u}ller}},
  \bibinfo{journal}{JHEP} \textbf{\bibinfo{volume}{01}}, \bibinfo{pages}{046}
  (\bibinfo{year}{2021}).

\bibitem[{\citenamefont{Grandchamp et~al.}(2006)\citenamefont{Grandchamp,
  Lumpkins, Sun, van Hees, and Rapp}}]{Grandchamp:2005yw}
\bibinfo{author}{\bibfnamefont{L.}~\bibnamefont{Grandchamp}},
  \bibinfo{author}{\bibfnamefont{S.}~\bibnamefont{Lumpkins}},
  \bibinfo{author}{\bibfnamefont{D.}~\bibnamefont{Sun}},
  \bibinfo{author}{\bibfnamefont{H.}~\bibnamefont{van Hees}}, \bibnamefont{and}
  \bibinfo{author}{\bibfnamefont{R.}~\bibnamefont{Rapp}},
  \bibinfo{journal}{Phys. Rev. C} \textbf{\bibinfo{volume}{73}},
  \bibinfo{pages}{064906} (\bibinfo{year}{2006}).

\bibitem[{\citenamefont{Andronic et~al.}(2024)}]{Andronic:2024oxz}
\bibinfo{author}{\bibfnamefont{A.}~\bibnamefont{Andronic}}
  \bibnamefont{et~al.}, \bibinfo{journal}{Eur. Phys. J. A}
  \textbf{\bibinfo{volume}{60}}, \bibinfo{pages}{88} (\bibinfo{year}{2024}).

\bibitem[{\citenamefont{Bala~et al.}(2022)}]{Bala:2021fkm}
\bibinfo{author}{\bibfnamefont{D.}~\bibnamefont{Bala~et al.}}
  (\bibinfo{collaboration}{HotQCD}), \bibinfo{journal}{Phys. Rev. D}
  \textbf{\bibinfo{volume}{105}}, \bibinfo{pages}{054513}
  (\bibinfo{year}{2022}).

\bibitem[{\citenamefont{Bazavov~et al.}(2024)}]{Bazavov:2023dci}
\bibinfo{author}{\bibfnamefont{A.}~\bibnamefont{Bazavov~et al.}}
  (\bibinfo{collaboration}{HotQCD}), \bibinfo{journal}{Phys. Rev. D}
  \textbf{\bibinfo{volume}{109}}, \bibinfo{pages}{074504}
  (\bibinfo{year}{2024}).

\bibitem[{\citenamefont{Ali et~al.}(2025)\citenamefont{Ali, Bala, Kaczmarek,
  and Pavan}}]{Ali:2025iux}
\bibinfo{author}{\bibfnamefont{S.}~\bibnamefont{Ali}},
  \bibinfo{author}{\bibfnamefont{D.}~\bibnamefont{Bala}},
  \bibinfo{author}{\bibfnamefont{O.}~\bibnamefont{Kaczmarek}},
  \bibnamefont{and} \bibinfo{author}{\bibnamefont{Pavan}}
  (\bibinfo{year}{2025}), \eprint{2505.11313}.

\bibitem[{\citenamefont{Mannarelli and Rapp}(2005)}]{Mannarelli:2005pz}
\bibinfo{author}{\bibfnamefont{M.}~\bibnamefont{Mannarelli}} \bibnamefont{and}
  \bibinfo{author}{\bibfnamefont{R.}~\bibnamefont{Rapp}},
  \bibinfo{journal}{Phys. Rev. C} \textbf{\bibinfo{volume}{72}},
  \bibinfo{pages}{064905} (\bibinfo{year}{2005}).

\bibitem[{\citenamefont{Cabrera and Rapp}(2007)}]{Cabrera:2006wh}
\bibinfo{author}{\bibfnamefont{D.}~\bibnamefont{Cabrera}} \bibnamefont{and}
  \bibinfo{author}{\bibfnamefont{R.}~\bibnamefont{Rapp}},
  \bibinfo{journal}{Phys. Rev. D} \textbf{\bibinfo{volume}{76}},
  \bibinfo{pages}{114506} (\bibinfo{year}{2007}).

\bibitem[{\citenamefont{van Hees et~al.}(2008)\citenamefont{van Hees,
  Mannarelli, Greco, and Rapp}}]{vanHees:2007me}
\bibinfo{author}{\bibfnamefont{H.}~\bibnamefont{van Hees}},
  \bibinfo{author}{\bibfnamefont{M.}~\bibnamefont{Mannarelli}},
  \bibinfo{author}{\bibfnamefont{V.}~\bibnamefont{Greco}}, \bibnamefont{and}
  \bibinfo{author}{\bibfnamefont{R.}~\bibnamefont{Rapp}},
  \bibinfo{journal}{Phys. Rev. Lett.} \textbf{\bibinfo{volume}{100}},
  \bibinfo{pages}{192301} (\bibinfo{year}{2008}).

\bibitem[{\citenamefont{Riek and Rapp}(2010)}]{Riek:2010fk}
\bibinfo{author}{\bibfnamefont{F.}~\bibnamefont{Riek}} \bibnamefont{and}
  \bibinfo{author}{\bibfnamefont{R.}~\bibnamefont{Rapp}},
  \bibinfo{journal}{Phys. Rev. C} \textbf{\bibinfo{volume}{82}},
  \bibinfo{pages}{035201} (\bibinfo{year}{2010}).

\bibitem[{\citenamefont{Liu and Rapp}(2018)}]{Liu:2017qah}
\bibinfo{author}{\bibfnamefont{S.~Y.~F.} \bibnamefont{Liu}} \bibnamefont{and}
  \bibinfo{author}{\bibfnamefont{R.}~\bibnamefont{Rapp}},
  \bibinfo{journal}{Phys. Rev. C} \textbf{\bibinfo{volume}{97}},
  \bibinfo{pages}{034918} (\bibinfo{year}{2018}).

\bibitem[{\citenamefont{Blankenbecler and Sugar}(1966)}]{BbS1966}
\bibinfo{author}{\bibfnamefont{R.}~\bibnamefont{Blankenbecler}}
  \bibnamefont{and} \bibinfo{author}{\bibfnamefont{R.}~\bibnamefont{Sugar}},
  \bibinfo{journal}{Phys. Rev.} \textbf{\bibinfo{volume}{142}},
  \bibinfo{pages}{1051} (\bibinfo{year}{1966}).

\bibitem[{\citenamefont{Thompson}(1970)}]{Thompson1970}
\bibinfo{author}{\bibfnamefont{R.~H.} \bibnamefont{Thompson}},
  \bibinfo{journal}{Phys. Rev. D} \textbf{\bibinfo{volume}{1}},
  \bibinfo{pages}{110} (\bibinfo{year}{1970}).

\bibitem[{\citenamefont{He et~al.}(2023)\citenamefont{He, van Hees, and
  Rapp}}]{He_2023}
\bibinfo{author}{\bibfnamefont{M.}~\bibnamefont{He}},
  \bibinfo{author}{\bibfnamefont{H.}~\bibnamefont{van Hees}}, \bibnamefont{and}
  \bibinfo{author}{\bibfnamefont{R.}~\bibnamefont{Rapp}},
  \bibinfo{journal}{Prog. Part. Nucl. Phys} \textbf{\bibinfo{volume}{130}},
  \bibinfo{pages}{104020} (\bibinfo{year}{2023}).

\bibitem[{\citenamefont{Wu et~al.}(2025)\citenamefont{Wu, Tang, and
  Rapp}}]{Wu:2025hlf}
\bibinfo{author}{\bibfnamefont{B.}~\bibnamefont{Wu}},
  \bibinfo{author}{\bibfnamefont{Z.}~\bibnamefont{Tang}}, \bibnamefont{and}
  \bibinfo{author}{\bibfnamefont{R.}~\bibnamefont{Rapp}}
  (\bibinfo{year}{2025}), \eprint{2503.10089}.

\bibitem[{\citenamefont{Haftel and Tabakin}(1970)}]{Haftel:1970zz}
\bibinfo{author}{\bibfnamefont{M.~I.} \bibnamefont{Haftel}} \bibnamefont{and}
  \bibinfo{author}{\bibfnamefont{F.}~\bibnamefont{Tabakin}},
  \bibinfo{journal}{Nucl. Phys. A} \textbf{\bibinfo{volume}{158}},
  \bibinfo{pages}{1} (\bibinfo{year}{1970}).

\bibitem[{\citenamefont{Holz and Glockle}(1988)}]{HOLZ1988131}
\bibinfo{author}{\bibfnamefont{J.}~\bibnamefont{Holz}} \bibnamefont{and}
  \bibinfo{author}{\bibfnamefont{W.}~\bibnamefont{Glockle}},
  \bibinfo{journal}{J. Comput. Phys} \textbf{\bibinfo{volume}{76}},
  \bibinfo{pages}{131} (\bibinfo{year}{1988}).

\bibitem[{\citenamefont{Rice and Kim}(1993)}]{Rice1993}
\bibinfo{author}{\bibfnamefont{R.~A.} \bibnamefont{Rice}} \bibnamefont{and}
  \bibinfo{author}{\bibfnamefont{Y.~E.} \bibnamefont{Kim}},
  \bibinfo{journal}{Few-Body Syst.} \textbf{\bibinfo{volume}{14}},
  \bibinfo{pages}{127} (\bibinfo{year}{1993}).

\bibitem[{\citenamefont{Elster et~al.}(1998)\citenamefont{Elster, Thomas, and
  Glöckle}}]{Elster_1998}
\bibinfo{author}{\bibfnamefont{C.}~\bibnamefont{Elster}},
  \bibinfo{author}{\bibfnamefont{J.~H.} \bibnamefont{Thomas}},
  \bibnamefont{and} \bibinfo{author}{\bibfnamefont{W.}~\bibnamefont{Glöckle}},
  \bibinfo{journal}{Few-Body Syst.} \textbf{\bibinfo{volume}{24}},
  \bibinfo{pages}{55–79} (\bibinfo{year}{1998}).

\bibitem[{\citenamefont{Karsch et~al.}(1988)\citenamefont{Karsch, Mehr, and
  Satz}}]{Karsch1988ER}
\bibinfo{author}{\bibfnamefont{F.}~\bibnamefont{Karsch}},
  \bibinfo{author}{\bibfnamefont{M.~T.} \bibnamefont{Mehr}}, \bibnamefont{and}
  \bibinfo{author}{\bibfnamefont{H.}~\bibnamefont{Satz}}, \bibinfo{journal}{Z.
  Phys. C - Particles and Fields} \textbf{\bibinfo{volume}{37}},
  \bibinfo{pages}{617} (\bibinfo{year}{1988}).

\bibitem[{\citenamefont{Bala and Datta}(2020)}]{Bala_2020}
\bibinfo{author}{\bibfnamefont{D.}~\bibnamefont{Bala}} \bibnamefont{and}
  \bibinfo{author}{\bibfnamefont{S.}~\bibnamefont{Datta}},
  \bibinfo{journal}{Phys. Rev. D} \textbf{\bibinfo{volume}{101}}
  (\bibinfo{year}{2020}).

\bibitem[{\citenamefont{Caia et~al.}(2004)\citenamefont{Caia, Pascalutsa, and
  Wright}}]{Caia:2003ke}
\bibinfo{author}{\bibfnamefont{G.}~\bibnamefont{Caia}},
  \bibinfo{author}{\bibfnamefont{V.}~\bibnamefont{Pascalutsa}},
  \bibnamefont{and} \bibinfo{author}{\bibfnamefont{L.~E.}
  \bibnamefont{Wright}}, \bibinfo{journal}{Phys. Rev. C}
  \textbf{\bibinfo{volume}{69}}, \bibinfo{pages}{034003}
  (\bibinfo{year}{2004}).

\end{thebibliography}
\end{document}